\documentclass[pdflatex,sn-mathphys-num]{sn-jnl}

\usepackage[utf8]{inputenc}
\usepackage{graphicx}
\usepackage{amsmath,amssymb,amsfonts,amsthm}
\usepackage{booktabs}
\usepackage{multirow}
\usepackage{tabularx}
\usepackage{xcolor}
\usepackage{hyperref}
\usepackage{float}
\usepackage{enumitem}

\usepackage{listings}

\lstdefinelanguage{json}{
    basicstyle=\normalfont\ttfamily,
    showstringspaces=false,
    breaklines=true,
    frame=lines,
    backgroundcolor=\color{white},
    literate=
     *{0}{{{\color{blue}0}}}{1}
      {1}{{{\color{blue}1}}}{1}
      {2}{{{\color{blue}2}}}{1}
      {3}{{{\color{blue}3}}}{1}
      {4}{{{\color{blue}4}}}{1}
      {5}{{{\color{blue}5}}}{1}
      {6}{{{\color{blue}6}}}{1}
      {7}{{{\color{blue}7}}}{1}
      {8}{{{\color{blue}8}}}{1}
      {9}{{{\color{blue}9}}}{1}
      {:}{{{\color{red}{:}}}}{1}
      {,}{{{\color{red}{,}}}}{1}
      {\{}{{{\color{violet}{\{}}}}{1}
      {\}}{{{\color{violet}{\}}}}}{1}
      {[}{{{\color{violet}{[}}}}{1}
      {]}{{{\color{violet}{]}}}}{1},
    morestring=[b]",
    stringstyle=\color{green!50!black},
}

\usepackage{tikz}
\usetikzlibrary{positioning,arrows.meta,calc}
\usepackage{placeins}
\usepackage{adjustbox}

\usepackage{colortbl}

\usepackage{xurl}
\usepackage{array}
\usepackage{ragged2e}
\begin{document}
\title{Toward Standardized Quantum Provenance: A Cross-Provider Analysis, Unified API, and Reference Prototype}

\author*[1,2]{\fnm{Jouni} \sur{Peltonen}}\email{jouni.peltonen@jyu.fi}
\author[1]{\fnm{Vlad} \sur{Stirbu}}\email{vlad.a.stirbu@jyu.fi}
\author[1]{\fnm{Tommi} \sur{Mikkonen}}\email{tommi.j.mikkonen@jyu.fi}
\author[3]{\fnm{Cesare} \sur{Pautasso}}\email{cesare.pautasso@usi.ch}

\affil[1]{\orgname{University of Jyväskylä}, \orgaddress{\city{Jyväskylä}, \country{Finland}}}
\affil[2]{\orgname{QMill Oy}, \orgaddress{\city{Espoo}, \country{Finland}}}
\affil[3]{\orgname{Università della Svizzera Italiana}, \orgaddress{\city{Lugano}, \country{Switzerland}}}

\maketitle

\begin{abstract}

Quantum software development requires provenance describing programs,
compilation, execution, hardware characterization, results, and software
environments, but providers expose this information through heterogeneous
software development kits, application programming interfaces, and resource
models. We analyze publicly documented provenance access across 15 quantum
platforms spanning five hardware technologies and find fragmented, incomplete
coverage, with compilation provenance weakest. We propose an evidence-aware
OpenAPI~3.1 provenance contract and provider-adapter architecture, evaluated
through a fixture-backed reference prototype at QMill covering Amazon Braket,
IBM Quantum, and IonQ. All records validate against one common contract while
preserving provider-specific semantics, explicit evidence origins, and
graceful handling of incomplete data.
\end{abstract}


\section{Introduction}
\label{sec:introduction}

Quantum computers have the potential to support advances in areas such as
materials science, chemistry, optimization, and finance. Recent achievements,
including Google's demonstration of quantum error correction below the
surface-code threshold~\cite{Acharya2024} and IBM's utility-oriented
experiments~\cite{Kim2023}, indicate progress from small proof-of-concept
demonstrations toward larger and more sustained quantum computations.

Current noisy intermediate-scale quantum (NISQ) devices nevertheless remain
constrained by noise, limited connectivity, finite coherence times, and
time-varying hardware performance~\cite{Preskill_2018}. Developing and
evaluating quantum algorithms on these devices requires knowledge of more than
the algorithm and its final result. Relevant context includes the submitted
and compiled program representations, target-device properties,
characterization data, compilation settings, execution history, result
transformations, and the software environment used to prepare and retrieve the
execution.

This contextual information is collectively referred to as
\emph{provenance data}. Provenance records the origin, transformations,
execution conditions, and interpretation context of a computational result. In
quantum software development, it supports analysis of why an execution
produced a particular result, comparison of algorithm behavior across devices
and time periods, investigation of compilation and hardware variation, and
reproduction of experiments as closely as the underlying platforms permit.

Quantum provenance presents particular challenges because quantum hardware is
typically remote, provider-managed, probabilistic, and time-dependent. The same
program submitted at two different times may be compiled differently or
executed under different device conditions. Some relevant information is
available only from the provider, whereas other information, such as the local
compiler configuration or software environment, may exist only in the
submitting application. A useful provenance record must therefore combine
provider and application context while preserving the source, temporal
relevance, and evidential strength of each value.

The quantum-computing ecosystem currently lacks a standardized mechanism for
retrieving such records. Hardware providers and cloud platforms expose
provenance through heterogeneous software development kits (SDKs), 
representational state transfer (REST) application programming interfaces (APIs),
generated clients, gRPC services, and cloud resource models. Their interfaces
use inconsistent terminology, distribute related information across different
resources, and provide varying access to programs, characterization,
compilation, execution, results, and software metadata. An absent field may
mean that the provider does not expose it, the application did not capture it,
the field is inapplicable to the technology, or the relevant resource is no
longer available. A normalized interface must distinguish these cases rather
than treating them as equivalent missing values.

This fragmentation creates a provenance-integration problem in addition to the
integration required to execute quantum programs. Multi-provider applications
must implement provider-specific retrieval, normalization, error handling, and
compatibility logic. They must also avoid misleading transformations, such as
presenting current device properties as the characterization used during an
earlier execution or treating locally captured compilation settings as
provider-returned information. These concerns affect both scientific
reproducibility and the maintainability of quantum software.

As quantum computing matures, established software-engineering practices such
as stable interfaces, machine-readable contracts, explicit semantics,
versioning, and automated conformance testing become increasingly important.
Recent initiatives on unified quantum interfaces, such as the Quantum Intermediate 
Representation (QIR)~\cite{qir}, have focused primarily on program representation, 
compilation, and execution portability. Provenance access has received
less attention, although standardized program representations alone do not identify
which representation was executed, how it was transformed, what device
conditions were available, or how the result was obtained.

This paper addresses the problem through a design-science research approach
combining comparative analysis, artifact design, iterative refinement, and
reference-prototype evaluation. We analyze the publicly documented
programmatic interfaces of 15 quantum platforms: 12 hardware-provider
platforms and three cloud aggregators. The included platforms span
superconducting, trapped-ion, neutral-atom, photonic, and quantum-annealing
hardware technologies. The evidence consists of public SDK and API
documentation, schemas, generated models, source code, and examples; the study
does not require authenticated provider access.

Building on the Quantum Provenance (QProv) model~\cite{weder2021qprov}, we
examine how its provenance concepts are exposed through heterogeneous provider
interfaces and extend the model where cross-provider access requires
additional context. The resulting design includes software and retrieval
context, evidence and availability qualifiers, artifact roles, temporal
characterization associations, technology-specific extensions, and separate
identification of the platform managing an execution and the provider
operating the hardware.

Based on the gaps identified in the provider comparison, we define a unified,
evidence-aware provenance API using OpenAPI~3.1~\cite{OpenAPI}. The API normalizes the
structure of provenance records while retaining differences in origin,
semantics, temporal relevance, and availability. A provider-adapter pattern
allows existing provider interfaces to be integrated incrementally without
requiring coordinated adoption of the proposed API by every provider.

We evaluate the design through a fixture-backed reference prototype integrated
with QMill's provider-executor architecture. The principal demonstration
compares an Amazon Braket execution routed to IonQ hardware with an IBM Quantum
execution accessed through a direct provider interface. A direct IonQ mapping
provides an additional case with a different result representation. The
prototype uses deterministic, sanitized fixtures so that its evaluation is
reproducible without provider accounts, credentials, or network access.

The three representative completed-job records validate against the same
reduced contract. A fixed inventory classifies 21 provenance attributes in
each case, and no completed record contains an unclassified inventory
attribute. Additional experiments demonstrate schema-valid partial records,
preservation of provider-specific result semantics, localization of a
provider-input change to the affected adapter, and backward-compatible
additive evolution of the common contract.

\noindent\textbf{\textit{Contributions.}}
This paper makes the following contributions:

\begin{itemize}

    \item \textbf{Cross-provider empirical analysis.}
    A systematic comparison of publicly documented provenance access across 15
    quantum platforms spanning five hardware technologies. The analysis
    identifies fragmented access mechanisms, inconsistent terminology,
    incomplete category coverage, and particularly limited compilation
    provenance (Section~\ref{subsec:provdataexposed}).

    \item \textbf{Evidence-aware provenance model.}
    An extension of QProv with software and access context, evidence and
    availability qualifiers, artifact roles, characterization associations,
    platform-versus-hardware-provider attribution, and support for
    technology-specific extensions
    (Sections~\ref{subsec:keydata} and~\ref{sec:apidesign}).

    \item \textbf{Unified provenance API and adapter architecture.}
    A provider-independent API defined through an OpenAPI~3.1 contract and explicit 
    design principles, together with an adapter pattern that permits incremental 
    integration of heterogeneous provider interfaces (Section~\ref{sec:apidesign}).

    \item \textbf{Reference prototype and industrial demonstration.}
    A fixture-backed implementation in QMill's provider-executor architecture
    for Amazon Braket, IBM Quantum, and direct IonQ, accompanied by a fixed
    provenance-attribute inventory and reproducible evaluation artifacts
    (Section~\ref{sec:casestudy}).

    \item \textbf{Evaluation of incomplete data and interface evolution.}
    Experiments evaluating partial provenance records, preservation of
    provider-specific semantics, isolation of provider-input changes, and
    backward-compatible additive contract evolution
    (Sections~\ref{subsec:prototype-results} and~\ref{sec:discussion}).

\end{itemize}

The remainder of this paper is structured as follows.
Section~\ref{sec:background} reviews quantum-computing fundamentals, quantum
software-development practices, and provenance concepts.
Section~\ref{sec:methodology} describes the research methodology and evidence
collection.
Section~\ref{sec:design} identifies the key provenance attributes and analyzes
their documented exposure across the included platforms.
Section~\ref{sec:apidesign} presents the unified API, design principles,
provider-adapter pattern, and adoption strategy.
Section~\ref{sec:casestudy} describes the QMill reference prototype and its
evaluation.
Section~\ref{sec:discussion} answers the research questions, interprets the
results, discusses ecosystem implications, and examines threats to validity.
Finally, Section~\ref{conclusions} summarizes the findings and outlines future
work.

\section{Background and related work}
\label{sec:background}

\subsection{Quantum computers and their calibration}
Implementing a gate-based quantum algorithm on current NISQ devices presents several challenges arising from their inherent noise and limited quantum resources~\cite{Leymann_2020}. Qubits are highly susceptible to decoherence and other forms of quantum noise, which can introduce errors into quantum computations~\cite{nielsen2010quantum}. Decoherence refers to the loss of quantum coherence, during which qubits gradually lose their quantum properties and begin to behave like classical bits. This can occur due to interactions with the environment, such as thermal fluctuations, electromagnetic interference, and other sources of noise.

To mitigate such errors, calibrating quantum circuits is essential. Calibration data are time-sensitive; qubit performance can drift significantly even within hours, making timestamped calibration snapshots essential for reproducibility. Calibration involves adjusting the parameters of the quantum gates and qubits to ensure that they operate as intended. The process typically includes characterizing the performance of quantum gates, measuring error rates, and correcting for systematic biases. By doing so, one can improve the fidelity of quantum operations, which is a measure of the accuracy with which a quantum gate performs its intended function. Calibration data can then be used to improve results by utilising it as input to job execution, by taking into account the topology of the quantum computer or via readout error mitigation. 
An example of improved results achieved through these methods is demonstrated in~\cite{vttadvusage}.

\subsection{Quantum software development}

Quantum software development is an iterative process using multiple classical quantum simulators and actual quantum hardware backends~\cite{Kinanen2025,zhao2020landscapes}. Quantum computers are specialized hardware that cannot be accessed in the same way as classical computers. When an algorithm is executed on a quantum computer, the access always occurs through classical computers and associated software, using a specific SDK or API. SDKs abstract hardware details and provide high-level constructs, while APIs offer direct, language-agnostic access to quantum resources—typically via REST endpoints that SDKs internally consume. Quantum software engineering research has begun to systematize these practices, including development processes, testing, and tooling~\cite{zhao2020landscapes,piattini2020talavera,ali2022roadmap}, and empirical studies report that practitioners struggle with fragmented toolchains and rapidly evolving platform interfaces~\cite{destefano2022practices}. 

Quantum SDKs (e.g. Qiskit~\cite{qiskit}, Cirq~\cite{cirq}, Braket
\cite{braketsdk}) provide tools for writing, simulating, and executing quantum circuits. They typically include visualization, simulators, noise models, and optimizers. Quantum SDKs require familiarity with programming (usually Python). Quantum APIs allow users to submit jobs, query status, and retrieve results, often through a job management API. These are mainly used inside the SDK. 

\subsection{Quantum experiment tracking and provenance}

Reproducibility is a cornerstone of scientific software engineering~\cite{doi:10.1126/science.aah6168, peng2011reproducible}. In classical computing, standards like Docker containers, package managers, and CI/CD pipelines enable reproducible builds and executions, and artifact
evaluation and badging initiatives increasingly formalize reproducibility
expectations for published research~\cite{acmbadging}. Quantum computing currently lacks analogous infrastructure, with provenance fragmentation undermining reproducibility across providers.

Quantum program development typically starts with simulators and advances to physical quantum devices as algorithms mature. This progression requires systematic data collection, initially to verify algorithmic correctness, and later to capture hardware-dependent factors such as calibration parameters and qubit performance metrics. Given the experimental nature of current quantum hardware, execution data are essential not only for performance evaluation but also for reproducibility, since hardware reliability can vary between runs and simulators cannot fully replace real-device experimentation~\cite{10313593}.

Provenance refers to all data and meta-data describing the history of an object, such as a piece of digital data or a physical object~\cite{10.1007/s00778-017-0486-1}. Provenance can be categorized into several types, including workflow provenance~\cite{10.1145/1376616.1376772} and data provenance~\cite{10.1145/1084805.1084812}; Herschel et al.\ provide a broad survey of provenance concepts and systems~\cite{10.1007/s00778-017-0486-1}. Classical scientific-workflow systems such as Kepler, Taverna, and Pegasus demonstrated early that capturing execution provenance across heterogeneous computational resources requires explicit provenance models and adapters~\cite{ludascher2006kepler,deelman2015pegasus}. In quantum computing research, both are essential: workflow provenance ensures reproducibility across platforms, while data provenance ensures transparency in how raw quantum measurement data become interpretable results. Without provenance, results cannot be trusted or replicated, as device-specific variations, such as differing qubit fidelities, remain unknown.

The provenance data constitute the audit trail of how a quantum result was obtained~\cite{Leymann_2020}. Weder et al.~\cite{weder2021qprov} introduced the Quantum Provenance (QProv) model, defining core provenance data attributes for quantum computation and implementing a system that automatically captures and stores these attributes. Subsequent work integrated provenance into quantum workflow
modeling and orchestration~\cite{weder2020quantumworkflow}. Related to workflow provenance, Gamage et al.~\cite{gamage2025enhancingquantumsoftwaredevelopment} demonstrated how MLflow~\cite{mlflow}, a mature machine learning and artificial intelligence experiment-tracking framework, can similarly support quantum workflows, promoting reproducibility, collaboration, and integration with existing research ecosystems rather than requiring bespoke solutions. 

\section{Research methodology}
\label{sec:methodology}

\subsection{Design Science Research approach}

This study follows the Design Science Research (DSR) methodology~\cite{10.2753/MIS0742-1222240302}, a problem-centric approach for generating knowledge through the creation and evaluation of artifacts that address practical problems. DSR is particularly appropriate when research aims to produce prescriptive solutions—such as models, methods, or specifications—rather than solely descriptive analyses~\cite{10.2753/MIS0742-1222240302}. 

In DSR, research proceeds through iterative cycles of design, demonstration, and evaluation. The methodology emphasizes:
\begin{itemize}
    \item \textbf{Problem relevance:} Addressing a significant real-world challenge faced by practitioners
    \item \textbf{Design as artifact:} Producing tangible outputs (models, specifications, guidelines) that embody solutions
    \item \textbf{Evaluation rigor:} Systematically assessing artifact utility through analysis, case studies, or experimentation
    \item \textbf{Research contribution:} Advancing knowledge about the problem domain and solution approaches
\end{itemize}

The artifacts produced by this research are:
\begin{itemize}
    \item A \textbf{conceptual provenance schema} defining essential attributes
    for quantum experiment tracking and reproducibility
    \item A set of \textbf{design principles} for unified quantum provenance
    APIs
    \item A \textbf{formal API specification} with implementation guidance for
    incremental adoption
    \item A \textbf{fixture-backed reference prototype} implementing the
    specification for three provider integrations, together with a
    reproducible evaluation artifact comprising the OpenAPI contract, adapters,
    sanitized fixtures, attribute inventory, and generated evaluation report
\end{itemize}

\subsection{Research questions}

The overarching goal of this research is to reduce the complexity of accessing quantum provenance data in multi-provider environments. This goal is operationalized through four research questions:

\textbf{RQ1: What provenance attributes are essential for quantum algorithm development, execution tracking, and reproducibility?}

Addressing this question requires identifying the information necessary to fully reproduce and understand quantum experiments across diverse hardware platforms and workflow contexts.

\textbf{RQ2: To what extent do current quantum providers expose essential provenance data, and through what mechanisms?}

This question characterizes the current ecosystem state: which attributes are accessible, which remain hidden, and how access patterns (SDK vs. API, authentication requirements, data formats) vary across providers and technologies.

\textbf{RQ3: What design principles should guide the development of a unified quantum provenance API?}

This question addresses solution design: what criteria ensure that a unified
API is standardized yet extensible, semantically faithful yet
provider-independent, and pragmatic yet forward-compatible?

\textbf{RQ4: Can a unified, evidence-aware provenance interface be realized
over existing heterogeneous provider interfaces in a real-world quantum
software architecture, and how does it handle incomplete data and
provider-interface change?}

This question grounds the research in practical validation: it examines
whether the proposed contract can normalize structurally different provider
integrations within an industrial provider-executor architecture, and whether
missing data and provider-specific interface changes can be contained behind
a stable application-facing boundary.

\subsection{Solution objectives}

Based on the research questions, we define objectives that guide artifact design and provide evaluation criteria:

\begin{itemize}
   \item \textbf{O1 (Completeness):} The provenance schema must be able to
    represent the information necessary for reproducing quantum experiments
    across providers and technologies, including explicit representation of
    information that is unavailable, inapplicable, or captured outside the
    provider interface.
    \item \textbf{O2 (Interoperability):} The unified API specification must be implementable across diverse quantum technologies without imposing technology-specific constraints.
    \item \textbf{O3 (Accessibility):} The access mechanism must be language-agnostic, use standard protocols, and provide machine-readable schemas to minimize integration barriers.
    \item \textbf{O4 (Extensibility):} The design must accommodate future hardware capabilities and technology-specific parameters through well-defined extension mechanisms.
    \item \textbf{O5 (Pragmatism):} The specification must align with existing provider capabilities and API conventions to maximize adoption potential.
\end{itemize}

\subsection{Study design and data collection}
\label{subsec:study-design}

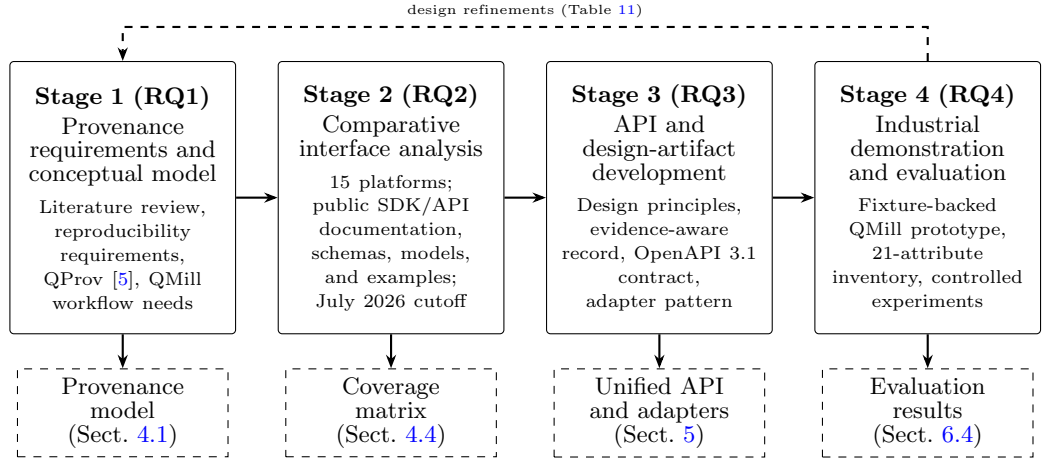
\begin{figure}[!htbp]
\centering
\begin{tikzpicture}[
    stage/.style={
        rectangle, draw, rounded corners=1.5pt,
        text width=2.75cm, minimum height=3.6cm,
        align=center, anchor=north, font=\scriptsize
    },
    output/.style={
        rectangle, draw, dashed,
        text width=2.55cm, minimum height=1.05cm,
        align=center, font=\scriptsize
    },
    flow/.style={-{Stealth[length=1.8mm]}, thick},
    feedback/.style={-{Stealth[length=1.8mm]}, thick, dashed}
]

\node[stage] (s1) at (0,0)
{\textbf{Stage 1 (RQ1)}\\[2pt]
Provenance requirements and conceptual model\\[3pt]
\footnotesize Literature review, reproducibility requirements,
QProv~\cite{weder2021qprov}, QMill workflow needs};

\node[stage, right=0.55cm of s1] (s2)
{\textbf{Stage 2 (RQ2)}\\[2pt]
Comparative interface analysis\\[3pt]
\footnotesize 15 platforms; public SDK/API documentation, schemas,
models, and examples; July 2026 cutoff};

\node[stage, right=0.55cm of s2] (s3)
{\textbf{Stage 3 (RQ3)}\\[2pt]
API and design-artifact development\\[3pt]
\footnotesize Design principles, evidence-aware record,
\mbox{OpenAPI~3.1} \mbox{contract}, adapter pattern};

\node[stage, right=0.55cm of s3] (s4)
{\textbf{Stage 4 (RQ4)}\\[2pt]
Industrial demonstration and evaluation\\[3pt]
\footnotesize Fixture-backed QMill prototype, 21-attribute inventory,
controlled experiments};

\node[output, below=0.45cm of s1] (o1)
{Provenance model\\(Sect.~\ref{subsec:keydata})};
\node[output, below=0.45cm of s2] (o2)
{Coverage \\ matrix\\(Sect.~\ref{subsec:provdataexposed})};
\node[output, below=0.45cm of s3] (o3)
{Unified API and adapters\\(Sect.~\ref{sec:apidesign})};
\node[output, below=0.45cm of s4] (o4)
{Evaluation results\\(Sect.~\ref{subsec:prototype-results})};

\draw[flow] (s1) -- (s2);
\draw[flow] (s2) -- (s3);
\draw[flow] (s3) -- (s4);
\draw[flow] (s1) -- (o1);
\draw[flow] (s2) -- (o2);
\draw[flow] (s3) -- (o3);
\draw[flow] (s4) -- (o4);

\draw[feedback] (s4.north) -- ++(0,0.45) -- 
    node[above, font=\tiny]{design refinements (Table~\ref{tab:dsriterations})}
    ($(s1.north)+(0,0.45)$) -- (s1.north);

\end{tikzpicture}
\caption{Four-stage study design. Solid arrows show the primary flow from
requirements to evaluation; each stage produces the artifact or evidence named
below it. The dashed feedback arrow indicates Design Science Research
build--evaluate cycles in which evaluation findings refined the provenance
model and API.}
\label{fig:studydesign}
\end{figure}

To address the research questions, we designed a four-stage study combining
literature analysis, systematic analysis of publicly documented provider
interfaces, iterative design-artifact development, and an industrial
reference-prototype demonstration. The stages correspond to the four research
questions while allowing findings from later evaluation activities to refine
the provenance model and API design. Figure~\ref{fig:studydesign} summarizes the 
four stages, their mapping to the research questions, and the artifact each stage produces.

\noindent\textbf{\textit{Stage 1: Provenance requirements and conceptual model
(RQ1).}}
The first stage identified the provenance required for quantum-algorithm
development, execution tracking, comparison, and reproducibility. The analysis combined a literature review, reproducibility requirements arising from remotely managed and time-varying hardware, the QProv model~\cite{weder2021qprov} as the principal conceptual reference, and practical requirements from QMill's multi-provider workflows. The resulting provenance model is presented in Section~\ref{subsec:keydata}; refinements arising from later evaluation activities are summarized in Section~\ref{subsec:design-refinement}.

\noindent\textbf{\textit{Stage 2: Platform selection and comparative interface
analysis (RQ2).}}
The second stage examined the documented programmatic exposure of provenance
across quantum hardware and cloud platforms. Platforms were selected using the
following criteria:

\begin{itemize}
    \item access to operational quantum hardware rather than only local
    simulation;

    \item a documented SDK, HTTP API, generated client, or equivalent
    programmatic interface;

    \item research or industrial relevance demonstrated through publications,
    maintained software, or an established user ecosystem; and

    \item architectural diversity across hardware technologies, direct-provider
    platforms, and cloud aggregators.
\end{itemize}

The final sample comprised 15 platforms: 12 hardware-provider platforms and
three cloud aggregators. For each platform, data collection examined official 
SDK, API, and service documentation; public OpenAPI specifications and generated 
client models where available; maintained SDK source code and package documentation; 
provider examples, tutorials, release notes, and public schemas; and peer-reviewed 
or technical literature clarifying provider terminology.

The study did not authenticate to deployed provider services or submit jobs
solely for the comparative analysis. An interface may therefore be described
as authenticated even though its capabilities were assessed through public
documentation and software artifacts. The analysis establishes documented
resource structures and operations but does not prove that every documented
field is populated for every account, backend, region, or completed job.

Exposed attributes were mapped to the provenance model using a three-level
coverage scale: comprehensive, partial, or unavailable. The classification
considered both whether an attribute category was documented and whether the
documented interface provided sufficient structure to retrieve and interpret
it. Publicly documented absence and insufficient public evidence were recorded
conservatively rather than inferred from analogous providers.

Access scope was evaluated separately from provenance completeness. A
restricted or invitation-only platform may document extensive provenance for
authorized users, whereas a broadly accessible platform may expose only
limited execution context. Authentication requirements were therefore treated
as interface-access properties rather than provenance-coverage ratings.

A cloud aggregator was defined as a platform that provides access to hardware
from multiple independent providers through a common account, job-management
model, and execution interface. A provider-neutral SDK was not classified as
an aggregator when its adapters connected separately to each provider's own
service.

Quantum-cloud interfaces evolve rapidly. The comparison is consequently a
time-bounded snapshot of named interfaces and software versions rather than a
permanent characterization of the platforms. Interface names, client versions,
source locations, and verification dates were recorded during data collection.
The final verification cutoff was July 2026. Previously available services
that had been retired, or current interfaces that could not be established
with sufficient public evidence, were excluded rather than assigned an
unavailable provenance rating.

\noindent\textbf{\textit{Stage 3: API and design-artifact development (RQ3).}}
The third stage synthesized the conceptual requirements and provider-analysis
findings into a unified provenance API. Recurring structures, semantic
differences, and accessibility gaps identified across provider interfaces
were combined with established API-design practices to derive design
principles, an evidence-aware provider-independent provenance record, a
reduced REST-compatible interface specified using OpenAPI~3.1, and a
provider-adapter pattern that separates provider-specific retrieval and
normalization from the application-facing contract. The evolving artifact was
evaluated through provider mappings, canonical examples, schema validation,
and controlled change scenarios.

Artifact development proceeded through documented build-evaluate cycles in
which evaluation activities exposed gaps in the evolving schema and API. The
gaps surfaced in each iteration and the resulting design decisions are
summarized retrospectively in Section~\ref{subsec:design-refinement}
(Table~\ref{tab:dsriterations}).

\noindent\textbf{\textit{Stage 4: Industrial demonstration and prototype
evaluation (RQ4).}}
The fourth stage assessed whether the proposed design can be realized within QMill's multi-provider architecture and how it behaves under incomplete data and provider-interface change. The industrial assessment included analysis of QMill's provider-executor architecture and integration responsibilities, qualitative assessment of provider-integration responsibilities in the code base, covering both execution and provenance concerns, and identification of provider-specific maintenance and compatibility concerns in multi-provider workflows.

The assessment was complemented by a fixture-backed reference prototype. The
prototype implements a reduced OpenAPI~3.1 contract, an in-process
provider-independent provenance service, a separate adapter registry, and
Braket, IBM, and direct-IonQ provenance mappings. Amazon Braket routed to IonQ
and IBM Quantum form the principal comparison between an aggregator-mediated
execution and a direct-provider execution. The direct-IonQ mapping provides an
additional result-semantics case.

The prototype evaluation used deterministic, sanitized fixtures rather than
live authenticated requests. The fixtures were constructed from the public
models and interface evidence examined in Stage 2. This approach allows the
normalization behavior and evaluation results to be reproduced without
credentials, network access, provider costs, or account-specific permissions.

The evaluation comprised:

\begin{itemize}
    \item validation of the three completed-job records against one common
    contract;

    \item classification of a fixed, versioned inventory of 21 provenance
    attributes for every provider case;

    \item measurement of handwritten provider-adapter and shared-implementation
    source lines;

    \item comparison of QMill-facing and adapter-side logical operations;

    \item a partial-record experiment in which characterization data were
    removed;

    \item a provider-input experiment in which an IBM-specific field was
    renamed; and

    \item an additive schema-evolution experiment evaluating backward
    compatibility and strict validation.
\end{itemize}

The committed canonical examples and evaluation report are generated from the
fixtures. The source code, OpenAPI contract,
attribute inventory, fixtures, generated records, and evaluation report are
provided in the accompanying research artifact~\cite{peltonen2026artifact}.

This four-stage design provides breadth through the comparison of multiple
platforms and hardware technologies, and depth through iterative artifact
development and an industrially situated, reproducible reference-prototype
evaluation.

\subsection{Evaluation approach}

We evaluate the research artifacts through multiple methods aligned with DSR
practice:

\textbf{Analytical evaluation.}
The comparative provider analysis assesses whether the proposed schema can
represent attributes beyond any single provider's current exposure, including
explicit representation of unavailable and inapplicable information,
validating completeness (O1). Technology diversity in the platform sample
(superconducting, trapped-ion, neutral-atom, photonic, and
quantum-annealing systems) validates interoperability (O2).

\textbf{Specification validation.}
The formal OpenAPI~3.1 specification demonstrates technical feasibility
through machine-readable, language-agnostic artifacts, validating
accessibility (O3). Namespaced extension mechanisms and an additive
schema-evolution experiment validate extensibility (O4).

\textbf{Prototype and experimental evaluation.}
A fixture-backed reference prototype in QMill's provider-executor
architecture validates that the contract can normalize an aggregator-mediated
execution, a direct-provider execution, and a differing result-semantics case
against one common schema. Controlled experiments evaluate behavior under
incomplete provenance, isolation of provider-input changes behind the adapter
boundary, and backward-compatible contract evolution. The industrial setting
and alignment with existing provider capabilities validate pragmatism (O5).

This multi-method evaluation combines observational evidence (documented
provider interfaces), analytical evidence (schema coverage and specification
validation), and experimental evidence (prototype validation and controlled
change scenarios) to triangulate artifact utility. The evaluation
intentionally excludes quantitative claims about integration- or
maintenance-effort reductions; establishing those would require an
independently implemented baseline and longitudinal observation of provider
interface changes (Section~\ref{sec:threats}).

\section{Provenance model and current-state analysis}
\label{sec:design}
The provenance schema defined in Section~\ref{subsec:keydata} serves a dual
role in this study: it is a design artifact in its own right, subsequently
refined into the unified API in Section~\ref{sec:apidesign}, and it provides
the analytical framework against which the documented provider interfaces are
compared in
Sections~\ref{subsec:access}-\ref{subsec:provdataexposed}.

\subsection{Definition of key quantum provenance data}
\label{subsec:keydata}

To reason about provenance consistently across providers, we require a schema that
links each execution to the software context in which it was produced and to the
calibration state of the device at that time. Rather than defining a new model, we
adopt the Quantum Provenance (QProv) model~\cite{weder2021qprov} as our conceptual
basis and refine it with a focused \emph{software-access context} needed for
cross-provider, API-based access. Figure~\ref{fig:qprovattrs} shows the resulting
schema: attributes inherited from or directly aligned with QProv are shown in
white, and the focused access-layer additions introduced in this work are shown in
green. The schema follows a one-record-per-job model that links what was executed,
where and how it was run, the device's calibration state at execution time, and
the resulting outputs.

\begin{figure*}[t]
\centering
\resizebox{\textwidth}{!}{%
\begin{tikzpicture}[
    category/.style={
        rectangle, draw=black, fill=white, line width=0.5pt,
        minimum height=0.7cm, align=center, font=\scriptsize\bfseries
    },
    attr/.style={
        rectangle, draw=black, fill=white, line width=0.4pt,
        minimum height=0.55cm, align=flush left, font=\scriptsize, text width=3.0cm
    },
    newattr/.style={ attr, fill=green!15, draw=green!50!black, line width=0.6pt }
]

\node[category, text width=16.8cm, minimum height=0.6cm] (title) at (0,0)
{Quantum provenance data};

\def\xA{-6.9}
\def\xB{-3.45}
\def\xC{0.0}
\def\xD{3.45}
\def\xE{6.9}

\node[category, text width=3.0cm] (qpuH)  at (\xA,-1.1) {Quantum computer / QPU};
\node[category, text width=3.0cm] (circH) at (\xB,-1.1) {Quantum circuit};
\node[category, text width=3.0cm] (execH) at (\xC,-1.1) {Execution};
\node[category, text width=3.0cm] (compH) at (\xD,-1.1) {Compilation};
\node[category, text width=3.0cm] (softH) at (\xE,-1.1) {Software and access context};

\node[attr] at (\xA,-2.0) {QPU name};
\node[attr] at (\xA,-2.6) {QPU version};
\node[attr] at (\xA,-3.2) {Last update};
\node[attr] at (\xA,-3.8) {Last calibration};
\node[attr] at (\xA,-4.4) {Queue size};
\node[attr] at (\xA,-5.0) {Max shots};
\node[attr] at (\xA,-5.6) {Is simulator};
\node[attr] at (\xA,-6.2) {Gate set};
\node[attr, minimum height=0.75cm] at (\xA,-6.95) {Qubits, connectivity, T1/T2, readout error};

\node[attr] at (\xB,-2.0) {Circuit name};
\node[attr] at (\xB,-2.6) {Circuit depth};
\node[attr] at (\xB,-3.2) {Circuit width};
\node[attr] at (\xB,-3.8) {Circuit size};
\node[attr] at (\xB,-4.4) {Circuit code URL};
\node[attr] at (\xB,-5.0) {Applied encoding};

\node[attr] at (\xC,-2.0) {Execution identifier};
\node[attr] at (\xC,-2.6) {Execution time};
\node[attr] at (\xC,-3.2) {Number of shots};
\node[attr, minimum height=0.75cm] at (\xC,-3.95) {Applied mitigation technique};
\node[attr] at (\xC,-4.7) {Circuit to execute};
\node[attr] at (\xC,-5.3) {Output data};
\node[attr] at (\xC,-5.9) {QPU};

\node[attr] at (\xD,-2.0) {Compilation time};
\node[attr] at (\xD,-2.6) {Optimization goal};
\node[attr] at (\xD,-3.2) {Random seed};
\node[attr] at (\xD,-3.8) {Compiler name};
\node[attr] at (\xD,-4.4) {Compiler provider};
\node[attr, minimum height=0.75cm] at (\xD,-5.15) {Compiler / transpiler version};
\node[attr] at (\xD,-6.0) {Input and output circuit};

\node[newattr] at (\xE,-2.0) {Access method};
\node[newattr, minimum height=0.75cm] at (\xE,-2.7) {Access method version};
\node[newattr] at (\xE,-3.4) {SDK/API version};
\node[newattr] at (\xE,-4.0) {Retrieval timestamp};
\node[newattr, minimum height=0.9cm] at (\xE,-5.05)
    {Data availability metadata\\{\tiny populated, missing, provider-limited}};
\node[newattr, minimum height=0.9cm] at (\xE,-6.52)
    {Technology-specific extensions\\{\tiny annealing, photonic, neutral atom}};

\node[category, text width=3.0cm] (provH) at (\xA,-8.2) {Provider};
\node[attr] at (\xA,-9.0) {Provider name};
\node[attr] at (\xA,-9.7) {Provider offering URL};

\node[attr, text width=4.4cm] at (\xC,-9.0)
    {\textbf{Legend:} white = aligned with QProv};
\node[newattr, text width=4.4cm] at (\xD,-9.0)
    {green = additions for cross-provider API provenance};

\end{tikzpicture}%
}
\caption{Quantum provenance schema used in this study. White attributes are
inherited from or directly aligned with QProv; green attributes represent focused
additions introduced for cross-provider API-based provenance access.}
\label{fig:qprovattrs}
\end{figure*}

The white attributes in Figure~\ref{fig:qprovattrs} are inherited from or directly
aligned with QProv and cover the circuit, quantum computer, compilation, and
execution categories, as well as provider identification. The focused additions of
this work, shown in green, form a dedicated \emph{software and access context}
category that QProv does not model. \textit{Access method} and \textit{access
method version} record the interface through which provenance was obtained (an
SDK, a REST API, or a cloud aggregator) since the same job can yield different
retrievable attributes depending on the access path. \textit{SDK/API version}
captures the client or library version, which materially affects transpilation and
execution results and is therefore essential for reproducibility. \textit{Retrieval
timestamp} distinguishes \emph{when} provenance was retrieved from when the
underlying computation or calibration occurred, preventing silent temporal
mismatches when calibration drifts between execution and data collection.
\textit{Data availability metadata} makes explicit which fields are populated,
missing, or provider-limited for a given job, so that gaps are declared rather than
silently absent. Finally, the \textit{technology-specific extensions} namespace
provides a well-defined container for parameters outside the gate model (such as
annealing offsets, trap geometry, or photonic squeezing levels), enabling
technology diversity without fragmenting the core schema.

\begin{table}[!htbp]
\centering
\caption{Schema design decisions relative to the QProv model: attributes adopted
directly, attributes mapped to existing QProv concepts, and the focused
access-layer additions introduced in this work. The complete attribute set and
its organization are shown in Figure~\ref{fig:qprovattrs}.}
\label{tab:qprovmapping}
\footnotesize
\setlength{\tabcolsep}{4pt}
\renewcommand{\arraystretch}{1.1}
\begin{tabularx}{\textwidth}{l l X}
\toprule
\textbf{Attribute} & \textbf{Design decision} & \textbf{Basis} \\
\midrule
Compiler / transpiler version & Adopt from QProv & Maps to the \texttt{compiler} class (\texttt{compilerVersion}); populated from the access layer. \\
Provider name / URL & Adopt from QProv & Maps to the \texttt{provider} class (\texttt{providerName}, \texttt{offeringUrl}). \\
Provider job identifier & Map to execution activity & Exposed as an external identifier for a QProv execution activity rather than a new core entity. \\
Job status & Runtime metadata & Reported as runtime state in the API response; not a core schema entity. \\
Access method / version & New (access layer) & QProv models provenance but not the interface used to retrieve it. \\
SDK/API version & New (access layer) & Client/library version affecting transpilation and results. \\
Retrieval timestamp & New (access layer) & Distinguishes \emph{when retrieved} from \emph{when computed or calibrated}. \\
Data availability metadata & New (access layer) & Declares populated, missing, or provider-limited fields. \\
Technology extensions namespace & New (access layer) & Optional container for annealing, photonic, and neutral-atom parameters. \\
\bottomrule
\end{tabularx}
\end{table}

As Table~\ref{tab:qprovmapping} summarizes, most attributes required for
cross-provider provenance are already modeled by QProv (including its
\emph{provider}, \emph{qpu}, \emph{compiler}, and \emph{execution} classes) so
our schema-level contribution is confined to the software-access context
described above.

Two attributes are treated specially rather than added as new core entities.
Provider-assigned job identifiers are exposed as external identifiers for QProv
execution activities; thus, rather than introducing a separate core job entity,
the unified API carries the provider job ID as part of execution metadata and maps
it to the corresponding execution activity. Job status is reported in the API
response as runtime state metadata, and is likewise not treated as a core element
of the provenance schema. Both remain valuable to algorithm developers: they
enable correlation of a reproducible provenance record with a provider's own
job-management interfaces for auditing, cost accounting, and retrieval of results
across long-running or queued executions.

Extensibility follows the same principle QProv uses for domain specialization: a
common core remains stable, while technology-dependent parameters (e.g., annealing
offsets, trap geometry, or photonic squeezing levels) are carried in an optional
\texttt{extensions} namespace rather than altering the core schema
(Section~\ref{subsec:apispec}).

\subsection{Programmatic access to quantum provenance data}
\label{subsec:access}

Quantum software development primarily relies on provider SDKs and client
libraries, most commonly distributed as Python packages~\cite{quantumsurvey}.
These libraries provide high-level abstractions for
circuit construction, compilation, job submission, status polling, result
retrieval, and device inspection. They reduce the amount of
provider-specific communication and serialization logic required from the
application. However, SDKs are typically tied to a particular language,
framework, provider, or version range, which limits their direct use from
heterogeneous software environments.

Provider SDKs commonly act as wrappers around lower-level service interfaces
rather than as independent access mechanisms. The underlying interfaces
include REST or REST-like HTTP APIs, gRPC services, generated cloud clients,
and provider-specific remote-procedure protocols. Some platforms combine
multiple mechanisms; for example, one interface may expose device and
calibration resources while another handles compilation, execution, and
result retrieval. Consequently, provenance accessibility cannot be inferred
only from whether a provider publishes a REST API. Relevant information may
be exposed through an SDK, a generated client, a service API, or a combination
of these mechanisms.

HTTP and gRPC service interfaces can support language-independent integration
more readily than a provider-specific Python SDK. They allow web
applications, dashboards, workflow systems, and provenance services to query
quantum resources without adopting the provider's primary programming
framework. REST-style interfaces are particularly suitable for persistent
resources such as jobs, devices, calibration records, sessions, and result
artifacts. gRPC and similar interfaces can support typed messages, streaming,
and efficient remote execution. Their practical accessibility nevertheless
depends on authentication, documentation, schema availability, and whether
the service is public, restricted, or available only through generated
clients.

Provenance data are also distributed across several resource types and
workflow stages. Submitted programs and execution settings may be stored in
job resources or external object storage; device properties and calibration
data may be exposed through separate endpoints; results may be returned
inline or through downloadable artifacts; and compilation information may
exist only in the client process that performed transpilation. A single job
query therefore rarely returns the complete provenance record. Reconstructing
an execution may require combining provider-supplied job, device,
calibration, and result data with application-captured circuit, compilation,
software-version, and retrieval context.

The analysis consequently evaluates provenance accessibility independently
of interface type. An attribute is considered provider-exposed when it can be
retrieved through a documented programmatic interface available to an
authorized user. Values available only in the originating application, local
compiler, or SDK process are classified as application-captured rather than
provider-supplied. This distinction informs both the IBM implementation
example in Section~\ref{subsec:ibm} and the cross-platform comparison in
Section~\ref{subsec:provdataexposed}.

\subsection{Illustrative attribute-level audit: IBM Quantum}
\label{subsec:ibm}

IBM Quantum is used to illustrate the attribute-level audit and mapping
procedure applied in the cross-provider analysis. The assessment was conducted
in July 2026 using Qiskit SDK~2.5.0~\cite{ibmsdk}, Qiskit IBM Runtime
0.47.0~\cite{ibmruntimeclient}, and Qiskit Runtime REST API version
2026-04-15~\cite{ibmapi}. It is based on the publicly documented client
methods, REST resources, schemas, and data models; field population in an
authenticated deployment was not operationally verified.

IBM's documented provenance is distributed across circuit objects, backend
configuration and properties, job details, metrics, inputs, and results.
Runtime client methods aggregate some of these resources. For example,
\texttt{RuntimeJobV2.metrics()} documents execution timestamps and QPU usage,
while job-associated backend properties and an optional
\texttt{calibration\_id} can provide temporal characterization context.
Compilation remains primarily application-side: transpiled circuits, layouts,
optimization settings, and random seeds are not systematically retained with
the completed job. Tables~\ref{tab:ibmcoverage-circuit}
and~\ref{tab:ibmcoverage-execution} present the resulting mapping and illustrate
the rating procedure used in Section~\ref{subsec:provdataexposed}.

\begin{table}[!htbp]
\centering
\caption{IBM Quantum provenance coverage (part 1 of 2): program, device, and compilation
attributes.}
\label{tab:ibmcoverage-circuit}
\footnotesize
\setlength{\tabcolsep}{3pt}
\renewcommand{\arraystretch}{0.86}

\begin{tabularx}{\textwidth}{p{2.4cm} X X}
\toprule
\textbf{Attribute} &
\textbf{SDK/client coverage} &
\textbf{REST API coverage} \\
\midrule

\multicolumn{3}{l}{\textit{Quantum program}} \\
\midrule

Used gates
& \cellcolor{gray!50}
  \texttt{circuit.data};
  \texttt{circuit.count\_ops()}
& \cellcolor{gray!25}
  Derivable from retained inputs; transpiled circuit unavailable \\

Used measurements
& \cellcolor{gray!50}
  \texttt{get\_instructions("measure")};
  \texttt{circuit.data}
& \cellcolor{gray!25}
  Derivable from retained inputs \\

Execution order
& \cellcolor{gray!50}
  Ordered \texttt{circuit.data}
& \cellcolor{gray!25}
  Submitted order reconstructable; transpiled order unavailable \\

Circuit width
& \cellcolor{gray!50}
  \texttt{width()};
  \texttt{num\_qubits}
& \cellcolor{gray!25}
  Derivable from retained inputs \\

Circuit depth
& \cellcolor{gray!50}
  \texttt{depth()}
& \cellcolor{gray!25}
  Derivable from retained inputs \\

Circuit size
& \cellcolor{gray!50}
  \texttt{size()}
& \cellcolor{gray!25}
  Derivable from retained inputs \\

Applied encoding
& \cellcolor{white}
  Requires application metadata
& \cellcolor{white}
  Only if included in job metadata \\

\midrule
\multicolumn{3}{l}{\textit{Quantum computer}} \\
\midrule

Number of qubits
& \cellcolor{gray!50}
  \texttt{backend.num\_qubits}
& \cellcolor{gray!50}
  Configuration \texttt{n\_qubits} \\

Decoherence times
& \cellcolor{gray!50}
  \texttt{properties()} or
  \texttt{qubit\_properties()}; T1/T2
& \cellcolor{gray!50}
  Backend properties; T1/T2 \\

Qubit connectivity
& \cellcolor{gray!50}
  \texttt{coupling\_map};
  \texttt{target}
& \cellcolor{gray!50}
  Configuration \texttt{coupling\_map} \\

Gate set
& \cellcolor{gray!50}
  \texttt{operation\_names},
  \texttt{basis\_gates}, or \texttt{target}
& \cellcolor{gray!50}
  Instructions and \texttt{basis\_gates} \\

Gate fidelities
& \cellcolor{gray!25}
  Derived as \(1-\textit{gate error}\)
& \cellcolor{gray!25}
  Gate error supplied; fidelity derived \\

Gate durations
& \cellcolor{gray!50}
  \texttt{target} or \texttt{properties()}
& \cellcolor{gray!50}
  Gate-length parameters \\

Readout fidelities
& \cellcolor{gray!25}
  Derived as \(1-\textit{readout error}\)
& \cellcolor{gray!25}
  Readout error supplied; fidelity derived \\

Provider identity
& \cellcolor{gray!25}
  Known from \texttt{QiskitRuntimeService}
& \cellcolor{gray!25}
  Implicit in service endpoint \\

Device identity
& \cellcolor{gray!50}
  \texttt{backend.name}
& \cellcolor{gray!50}
  Job \texttt{backend} or \texttt{backend\_name} \\

Job-associated properties
& \cellcolor{gray!25}
  \texttt{job.properties()} when available
& \cellcolor{gray!25}
  Retrieved using job calibration context \\

\midrule
\multicolumn{3}{l}{\textit{Compilation}} \\
\midrule

Qubit assignments
& \cellcolor{gray!50}
  \texttt{transpiled\_circuit.layout}
& \cellcolor{white}
  Transpiled layout not retained \\

Gate mappings
& \cellcolor{gray!50}
  Transpiled circuit and routing information
& \cellcolor{white}
  Client routing not exposed \\

Optimization goal
& \cellcolor{gray!25}
  \texttt{optimization\_level}; application-captured
& \cellcolor{white}
  Not systematically retained \\

Random seed
& \cellcolor{gray!25}
  \texttt{seed\_transpiler}; application-captured
& \cellcolor{white}
  Not systematically retained \\

Compilation time
& \cellcolor{gray!25}
  Local timing or pass-manager callbacks
& \cellcolor{white}
  Not exposed \\

Compiler version
& \cellcolor{gray!25}
  Local Qiskit and pass-manager versions
& \cellcolor{white}
  Runtime version may differ from client compiler \\

\bottomrule
\end{tabularx}

\footnotetext{\textbf{Legend:}
\protect\fcolorbox{black}{gray!50}{\strut\hspace{6pt}} direct/comprehensive;
\protect\fcolorbox{black}{gray!25}{\strut\hspace{6pt}} partial, derived, or conditional;
\protect\fcolorbox{black}{white}{\strut\hspace{6pt}} limited/unavailable.}
\end{table}

\begin{table}[!htbp]
\centering
\caption{IBM Quantum provenance coverage (part 2 of 2): execution and software-access
attributes.}
\label{tab:ibmcoverage-execution}
\footnotesize
\setlength{\tabcolsep}{3pt}
\renewcommand{\arraystretch}{0.84}

\begin{tabularx}{\textwidth}{p{2.4cm} X X}
\toprule
\textbf{Attribute} &
\textbf{SDK/client coverage} &
\textbf{REST API coverage} \\
\midrule

\multicolumn{3}{l}{\textit{Execution}} \\
\midrule

Input data
& \cellcolor{gray!50}
  \texttt{job.inputs}; primitive PUBs/options
& \cellcolor{gray!25}
  Job \texttt{params}, subject to retention and privacy \\

Output data
& \cellcolor{gray!50}
  \texttt{job.result()}
& \cellcolor{gray!50}
  \texttt{GET /v1/jobs/\{id\}/results} \\

Number of shots
& \cellcolor{gray!50}
  PUB, run options, or \texttt{job.inputs}
& \cellcolor{gray!25}
  Retained parameters; not necessarily top-level \\

Intermediate results
& \cellcolor{white}
  No general streaming support
& \cellcolor{white}
  Final results only \\

Number of iterations
& \cellcolor{gray!25}
  When retained in inputs or metadata
& \cellcolor{gray!25}
  When retained in job parameters \\

Execution timing
& \cellcolor{gray!50}
  \texttt{job.metrics()}; wall-clock and QPU usage
& \cellcolor{gray!50}
  Lifecycle timestamps and QPU timing \\

Error mitigation
& \cellcolor{gray!25}
  Requested mitigation or suppression options
& \cellcolor{gray!25}
  Options may be retained in \texttt{params} \\

Queue duration
& \cellcolor{gray!50}
  Derived from running minus created
& \cellcolor{gray!50}
  Derived from metrics timestamps \\

Completion time
& \cellcolor{gray!50}
  Metrics \texttt{finished}
& \cellcolor{gray!50}
  Metrics \texttt{finished} \\

Job identifier
& \cellcolor{gray!50}
  \texttt{job.job\_id()}
& \cellcolor{gray!50}
  Job \texttt{id} \\

Job status
& \cellcolor{gray!50}
  \texttt{job.status()}
& \cellcolor{gray!50}
  Job/state \texttt{status} \\

Calibration identifier
& \cellcolor{gray!25}
  Indirectly through job context
& \cellcolor{gray!25}
  \texttt{calibration\_id} when present \\

\midrule
\multicolumn{3}{l}{\textit{Software and access context}} \\
\midrule

Backend version
& \cellcolor{gray!50}
  \texttt{backend.backend\_version}
& \cellcolor{gray!50}
  Backend \texttt{backend\_version} \\

Qiskit version
& \cellcolor{gray!50}
  \texttt{qiskit.\_\_version\_\_}
& \cellcolor{gray!50}
  Metrics \texttt{qiskit\_version} \\

Runtime-client version
& \cellcolor{gray!50}
  Installed package version
& \cellcolor{gray!25}
  Caller data may omit exact version \\

REST API version
& \cellcolor{gray!25}
  Managed by Runtime client
& \cellcolor{gray!50}
  \texttt{IBM-API-Version} header \\

Runtime image
& \cellcolor{gray!50}
  \texttt{job.image}
& \cellcolor{gray!50}
  Job \texttt{runtime} when retained \\

Access method
& \cellcolor{gray!25}
  Known from the client used
& \cellcolor{gray!25}
  Known from direct REST access \\

Retrieval timestamp
& \cellcolor{gray!25}
  Application-captured
& \cellcolor{white}
  Not included in response \\

Availability metadata
& \cellcolor{white}
  Constructed by application
& \cellcolor{white}
  No field-level availability report \\

\bottomrule
\end{tabularx}

\footnotetext{The coverage legend is the same as in
Table~\ref{tab:ibmcoverage-circuit}.}
\end{table}

IBM illustrates that comparatively rich provider support does not by itself
produce one complete provenance record. Device and execution information are
well documented, but program, characterization, and result data must be
correlated across multiple resources. Detailed compilation and retrieval
context remain dependent on application-side capture. This fragmentation
motivates both the common record structure and the explicit evidence qualifiers
introduced in Section~\ref{sec:apidesign}.

\subsection{Comparative analysis of provider provenance data}
\label{subsec:provdataexposed}

To extend the findings from the detailed IBM analysis in
Section~\ref{subsec:ibm}, we examine a purposively selected and technologically
diverse set of quantum hardware and cloud platforms. The analysis compares how
their programmatic interfaces expose the provenance attributes defined in
Section~\ref{subsec:keydata}. The unit of analysis is the provider platform and
the documented SDK, client library, service API, or other programmatic interface
available during the data-collection period. Thus, platforms that share hardware
technology or client libraries may be analyzed separately when they are
independently operated and expose different authentication, job-management,
calibration, or data-retention interfaces.

For each platform, the exposed fields were mapped to the common provenance
categories. Individual attributes were classified according to whether they
were directly exposed, derivable from other exposed fields, capturable only by
the consuming application, available only through a particular access method,
restricted, unavailable, not applicable to the underlying technology, or not
verifiable from the available evidence. This distinction prevents
application-captured or inferred values from being presented as
provider-supplied provenance and avoids treating inapplicable gate-model
attributes as missing from non-gate-based systems.

The mapping provides a common basis for comparing provenance accessibility
despite differences in provider terminology, response structures, access
mechanisms, and quantum-computing paradigms. Provider-specific concepts (such
as backend properties, characterization records, calibration sets, processor
configurations, solver properties, device certificates, and working
graphs) were mapped to semantically corresponding attributes in the common
schema. Technology-specific attributes with no direct counterpart in the
common core were evaluated through the extensions mechanism rather than
automatically treated as missing.

\begin{table}[!htbp]
\centering
\caption{Criteria used to select quantum platforms for the comparative analysis.}
\label{tab:providercriteria}
\footnotesize
\setlength{\tabcolsep}{4pt}
\renewcommand{\arraystretch}{1.05}
\begin{tabularx}{\textwidth}{p{2.7cm} X}
\toprule
\textbf{Criterion} & \textbf{Operationalization} \\
\midrule

Active platform
& The platform offered access to physical quantum hardware, directly or through
a cloud service, during the data-collection period. Simulator-only platforms
were excluded. \\

Programmatic access
& The platform provided a documented SDK, client library, REST, gRPC, or other
service interface for job submission, status retrieval, result retrieval, or
device inspection. A public REST API was not required, because the
accessibility of provenance through different interface types is itself an
object of analysis. \\

Documented interface
& Sufficient official documentation, maintained source code, machine-readable
schemas, or authenticated interface information was available to identify and
evaluate provenance-related fields. Platforms that could not be assessed
beyond marketing material were excluded. \\

Ecosystem relevance
& The platform represented an active hardware or cloud ecosystem, as indicated
by maintained software integrations, current provider documentation, reported
research use, or availability through an established quantum cloud service. \\

Technology coverage
& The sample was selected to cover superconducting, trapped-ion, neutral-atom,
photonic, and quantum-annealing systems, including provenance attributes that
differ from those of the gate-based computing model. \\

Platform-type coverage
& Both hardware-provider platforms and cloud aggregators were included. The two
groups were analyzed separately because aggregators depend partly on provenance
supplied by their connected hardware providers. \\

\bottomrule
\end{tabularx}
\end{table}

The platforms included in the study were selected according to
Table~\ref{tab:providercriteria}. The sampling strategy was purposive rather
than exhaustive: its objective was to capture variation in hardware
technology, provider type, and programmatic access mechanism. The final sample
includes superconducting, trapped-ion, neutral-atom, photonic, and
quantum-annealing platforms, together with multi-provider cloud aggregators.

For cloud aggregators, the assessment distinguishes between metadata
consistently normalized by the aggregator and metadata merely passed through
from an underlying hardware provider. Rich metadata available for only one
connected device or through a direct provider adapter were not considered
evidence of comprehensive cross-provider coverage by the aggregator.

Because quantum cloud platforms evolve rapidly, the results represent a
time-bounded snapshot of the interfaces available at the July 2026 verification
cutoff. Provider capabilities were verified using official documentation,
release notes, maintained client implementations, machine-readable
specifications, and authenticated interface information or access where
available. SDK and API versions were recorded for each platform. Retired
services and interfaces that could not be verified were excluded rather than
classified as exposing limited provenance.

\begin{table}[!htbp]
\centering
\caption{Quantum hardware and cloud platforms included in the comparative
analysis. Interface names and access requirements reflect the public
documentation available during the data-collection period; authenticated 
interfaces were evaluated from public documentation and
models rather than through authenticated service requests.}
\label{tab:apilist}
\footnotesize
\setlength{\tabcolsep}{3pt}
\renewcommand{\arraystretch}{1.02}
\begin{tabularx}{\textwidth}{p{2.7cm} p{2.0cm} X X}
\toprule
\textbf{Platform} &
\textbf{Technology} &
\textbf{Primary SDK/client} &
\textbf{HTTP/API access} \\
\midrule

\multicolumn{4}{l}{\textit{Hardware-provider platforms}} \\
\midrule
AQT Arnica & Trapped ion & Qiskit AQT Provider & Authenticated AQT Public API  (OpenAPI 3.1) \\
D-Wave Leap & Quantum annealing & Ocean SDK & Authenticated Solver API (SAPI REST) \\
Google Quantum AI & Superconducting & Cirq & Restricted Quantum Engine API \\
IBM Quantum Platform & Superconducting & Qiskit Python SDK & Qiskit Runtime REST API \\
IonQ Quantum Cloud & Trapped ion & Qiskit IonQ Provider & Authenticated IonQ Quantum Cloud REST API \\
IQM Resonance & Superconducting & IQM Client (Qiskit and Cirq adapters) & Authenticated IQM Server REST API \\
Pasqal Cloud & Neutral atom & Pulser and pasqal-cloud & Authenticated Pasqal Cloud REST API (OpenAPI 3.0) \\
Quandela Cloud & Photonic & Perceval & Authenticated Quandela Cloud REST API \\
Quantinuum Nexus & Trapped ion & qnexus and pytket & Authenticated Nexus REST API (OpenAPI) \\
QuEra Aquila (Amazon Braket) & Neutral atom & Amazon Braket SDK & Authenticated Amazon Braket REST API \\
Rigetti QCS & Superconducting & pyQuil and QCS SDK & Authenticated OpenAPI HTTP and gRPC APIs \\
VTT QX & Superconducting & IQM Client and IQM Pulla & Authenticated VTT QX REST API \\
\midrule
\multicolumn{4}{l}{\textit{Cloud aggregators}} \\
\midrule
Amazon Braket & Multi-provider & Amazon Braket SDK & Authenticated Amazon Braket REST API \\
Microsoft Azure Quantum & Multi-provider & Microsoft QDK and Azure Quantum SDK & Authenticated Azure Quantum REST API (OpenAPI) \\
qBraid Cloud & Multi-provider & qBraid SDK & Authenticated qBraid Quantum Runtime REST API \\
\bottomrule
\end{tabularx}
\end{table}

The evidence sources for each platform are cited in the supplementary platform profiles~\cite{peltonen2026artifact}.

\subsubsection{Scope and interpretation of platform comparisons}

For hardware providers accessed through cloud aggregators, such as QuEra
Aquila through Amazon Braket, the hardware-provider row evaluates
provider-specific program, device, and result metadata, whereas the aggregator
row evaluates the consistency and completeness of the cloud platform's
cross-provider interface.

Some platforms share hardware technology, client libraries, or server data
models while exposing different service-level interfaces. In particular, VTT
Q50 was co-developed by VTT and IQM, and VTT QX supports IQM Client, Qiskit,
Cirq, and IQM Pulla interfaces. Nevertheless, IQM Resonance and VTT QX are
analyzed separately because they are independently operated platforms with
different service endpoints, authentication mechanisms, device offerings,
calibration policies, job-management resources, and data-retention behavior.
The unit of analysis is therefore the provider platform and its observable
programmatic interface, rather than the QPU manufacturer alone.

Quandela Cloud was included as the photonic hardware platform because it
provided active, documented, authenticated QPU access through Perceval at the
July 2026 verification cutoff. Xanadu was not included because its former
public cloud service and Borealis hardware access were no longer available,
and a current documented programmatic interface to Xanadu hardware could not
be verified. Retired interfaces were not used to characterize current
provenance coverage.

Platforms limited to simulators, development-stage hardware without
programmatic access, research middleware without an associated execution
service, or undocumented private interfaces were outside the scope of the
comparison.

\subsubsection{Platform-specific findings}
\label{subsubsec:platform-findings}

The complete platform profiles (assessed interfaces and versions, evidence
sources, and field-level rating justifications for all 15 platforms) are
provided as supplementary material in the accompanying research
artifact~\cite{peltonen2026artifact}. The assessment concerns provenance
documented through publicly accessible SDK and API documentation, as described 
in Section~\ref{subsec:study-design} (Stage 2).

Three platform contrasts illustrate the range of documented provenance
behavior underlying the coverage matrix. First, Quantinuum Nexus is the
category-level exception for compilation provenance: compilation is a
first-class job resource that retains input and output circuits, applied
passes, and optionally intermediate circuits, whereas on most other platforms
compilation artifacts exist only in the client process and are not associated
with the completed job. Second, D-Wave demonstrates why technology-specific
extensions are necessary: its applicable provenance is defined by annealing
problems, solver properties, working graphs, and minor embeddings rather than
gate-model circuits and coherence metrics, so treating absent gate-model
fields as missing data would misrepresent the platform. Third, QuEra accessed
through Amazon Braket illustrates the aggregator relationship: the submitted
analog program and execution metadata are retained in Braket's common task
model, while hardware-specific characterization and compilation detail depend
on what the underlying provider supplies, motivating the separate
representation of the execution platform and the hardware operator.

For aggregators generally, the assessment distinguishes common platform fields
from underlying-provider information passed through without cross-provider
normalization. A comprehensive aggregator rating requires consistent exposure
across supported providers; rich metadata available for only one connected
device does not establish comprehensive aggregator-level coverage.

\textbf{Cross-platform interpretation.}
The platform profiles reinforce three findings from the coverage matrix.
First, job identity, status, timestamps, shots, and results are the most
consistently exposed provenance. Second, device characterization varies both
by hardware technology and by the strength of its association with a
particular execution: some platforms document immutable or selectable
characterization identifiers, while others provide only current or
nearest-available device properties. Third, compilation provenance is
predominantly client-side and is rarely retained as a complete provider job
history. Software versions, retrieval context, and explicit availability
metadata are likewise usually left to the consuming application.

\subsubsection{Comparative coverage results}

A summary of the analysis is presented in
Table~\ref{tab:apiprovtable} as a three-level qualitative coverage matrix
indicating comprehensive, partial, and limited support for each provenance
category. Shading represents the relative completeness of the exposed data,
with darker tones indicating broader coverage of the applicable attributes.
Attributes classified as not applicable to a particular computing paradigm
were excluded from the corresponding coverage judgment rather than treated as
missing.

The matrix reveals both common practices and substantial differences across
provider platforms. Program and execution provenance are generally the
strongest categories: many platforms retain the submitted program or an
executable representation and expose job identifiers, lifecycle states,
timestamps, shot or run counts, and results. QPU provenance varies more
substantially. Some provider-operated platforms expose timestamped
characterizations, calibration sets, or execution-associated device snapshots,
whereas cloud aggregators generally provide less consistent calibration
coverage across their connected hardware providers. Software and access
context also remains incomplete because exact client-library versions,
retrieval timestamps, and local execution environments usually need to be
recorded by the consuming application.

None of the analyzed platforms exposes every key provenance attribute through
a single SDK or service interface. Compilation provenance is incomplete for
most platforms: logical-to-physical mappings, optimization passes, random
seeds, intermediate representations, compiler versions, and compilation
durations are often computed or retained only on the client side and are not
systematically associated with the completed job. Quantinuum Nexus is a
notable category-level exception, retaining compilation inputs, outputs, pass
information, and optionally intermediate circuits. Nevertheless, some fields,
including compiler-version and random-seed information, remain unavailable or
require application-side capture. Thus, a comprehensive category rating
indicates broad coverage of the applicable attributes rather than the
availability of every individual field.

The aggregator results provide an additional architectural insight. Amazon
Braket, Microsoft Azure Quantum, and qBraid expose relatively consistent
program-submission and execution abstractions, but their QPU and compilation
coverage depends partly on information supplied by the underlying hardware
providers. More standardized provenance interfaces at the hardware-provider
level would therefore simplify aggregator integration and improve the
consistency of provenance available to users. These findings form the
empirical basis for the discussion in
Section~\ref{sec:discussion}.

\begin{table}[t]
\centering
\caption{Comparative coverage of provenance data attributes across selected quantum providers}
\label{tab:apiprovtable}
\small
\begin{tabular}{lccccc}  
\toprule
\textbf{Provider} & \textbf{Circuit} & \textbf{Computer} & \textbf{Compilation} & \textbf{Execution} & \textbf{Software} \\
\midrule
\multicolumn{6}{l}{\textit{Hardware Providers}} \\
\midrule
AQT
& \cellcolor{gray!50}  
& \cellcolor{gray!75}  
& \cellcolor{gray!50}  
& \cellcolor{gray!50}  
& \cellcolor{gray!50}  
\\
D-Wave
& \cellcolor{gray!50}  
& \cellcolor{gray!75}  
& \cellcolor{gray!50}  
& \cellcolor{gray!75}  
& \cellcolor{gray!50}  
\\
Google Quantum AI
& \cellcolor{gray!75}  
& \cellcolor{gray!75}  
& \cellcolor{gray!50}  
& \cellcolor{gray!75}  
& \cellcolor{gray!50}  
\\
IBM         & \cellcolor{gray!50} & \cellcolor{gray!75} & \cellcolor{gray!50} & \cellcolor{gray!75} & \cellcolor{gray!75} \\
IonQ
& \cellcolor{gray!50}  
& \cellcolor{gray!75}  
& \cellcolor{gray!50}  
& \cellcolor{gray!75}  
& \cellcolor{gray!50}  
\\
IQM
& \cellcolor{gray!75}  
& \cellcolor{gray!75}  
& \cellcolor{gray!50}  
& \cellcolor{gray!75}  
& \cellcolor{gray!50}  
\\
Pasqal
& \cellcolor{gray!50}  
& \cellcolor{gray!50}  
& \cellcolor{gray!50}  
& \cellcolor{gray!75}  
& \cellcolor{gray!50}  
\\
Quandela
& \cellcolor{gray!50}  
& \cellcolor{gray!50}  
& \cellcolor{gray!50}  
& \cellcolor{gray!75}  
& \cellcolor{gray!50}  
\\
Quantinuum
& \cellcolor{gray!75}  
& \cellcolor{gray!50}  
& \cellcolor{gray!75}  
& \cellcolor{gray!75}  
& \cellcolor{gray!50}  
\\
QuEra
& \cellcolor{gray!75}  
& \cellcolor{gray!50}  
& \cellcolor{gray!50}  
& \cellcolor{gray!75}  
& \cellcolor{gray!50}  
\\
Rigetti
& \cellcolor{gray!50}  
& \cellcolor{gray!75}  
& \cellcolor{gray!50}  
& \cellcolor{gray!50}  
& \cellcolor{gray!50}  
\\
VTT
& \cellcolor{gray!75}  
& \cellcolor{gray!75}  
& \cellcolor{gray!50}  
& \cellcolor{gray!75}  
& \cellcolor{gray!50}  
\\
\midrule
\multicolumn{6}{l}{\textit{Cloud Aggregators}} \\
\midrule
Amazon Braket
& \cellcolor{gray!75}  
& \cellcolor{gray!50}  
& \cellcolor{gray!50}  
& \cellcolor{gray!75}  
& \cellcolor{gray!50}  
\\
Microsoft Azure Quantum
& \cellcolor{gray!75}  
& \cellcolor{gray!25}  
& \cellcolor{gray!50}  
& \cellcolor{gray!75}  
& \cellcolor{gray!50}  
\\
qBraid
& \cellcolor{gray!75}  
& \cellcolor{gray!50}  
& \cellcolor{gray!50}  
& \cellcolor{gray!75}  
& \cellcolor{gray!50}  
\\
\bottomrule
\end{tabular}
\medskip
\small
\noindent\textbf{Legend:} 
\colorbox{gray!75}{\strut\hspace{8pt}} comprehensive (broad coverage of applicable attributes); 
\colorbox{gray!50}{\strut\hspace{8pt}} partial (substantial coverage with notable gaps); 
\colorbox{gray!25}{\strut\hspace{8pt}} limited (few or no applicable attributes exposed)
\end{table}

\section{Unified provenance API design}
\label{sec:apidesign}

This section presents a unified provenance API that aggregates and qualifies
this information without replacing provider execution interfaces. A common
outward-facing contract is combined with provider adapters that retrieve data
from heterogeneous upstream services. The API normalizes record structure
while retaining provider-native identifiers, technology-specific semantics,
evidence origin, and explicit information about unavailable attributes. The
reference prototype in Section~\ref{sec:casestudy} implements and evaluates a
focused subset of this design.

\subsection{Design principles and criteria}
\label{subsec:designprinciples}

Drawing on the APIstic benchmark study by Serbout and
Pautasso~\cite{serbout2024apistic}, OpenAPI, established
API-design guidance~\cite{lauret2019design}, and the provider analysis, the
design is guided by three high-level criteria:

\begin{itemize}
    \item \textbf{Standardization and interoperability}: provenance should be
    available through a provider-independent but extensible schema.

    \item \textbf{Explicit, machine-readable semantics}: fields, types, units,
    evidence origin, availability, and schema versions should be formally
    specified.

    \item \textbf{Incremental provider integration}: adapters should map
    existing REST, gRPC, SDK, generated-client, and storage interfaces into the
    common contract without requiring providers to replace their execution
    services.
\end{itemize}

These criteria are operationalized through the following principles.

\noindent\textbf{\textit{P1: Schema standardization.}}
The API defines common names, structures, data types, units, and semantic
definitions for the provenance categories identified in
Section~\ref{subsec:keydata}. This reduces ambiguity caused by
provider-specific terminology and supports consistent storage and comparison.

\noindent\textbf{\textit{P2: Technology extensibility.}}
A stable common core is complemented by namespaced extensions for concepts
specific to gate-based, annealing, neutral-atom, photonic, analog, and other
quantum-computing technologies. Technology-specific programs and metrics are
not forced into superficially similar gate-model fields.

\noindent\textbf{\textit{P3: Artifact lineage.}}
The model distinguishes source, submitted, compiled, executed, environment,
and result artifacts. When available, artifacts identify their formats,
versions, producers, timestamps, hashes, storage references, and
transformation relationships.

\noindent\textbf{\textit{P4: Normalized and provider-native representations.}}
Normalized values retain provider-native identifiers and evidence references.
Provider-native records may be retained separately when permitted. For
aggregators, the model distinguishes normalized platform data from information
passed through from an underlying hardware provider.

\noindent\textbf{\textit{P5: HTTP-based outward interface.}}
The API exposes a resource-oriented HTTP interface while allowing adapters to
use any required upstream mechanism. Applications receive one consistent
interface even when adapters rely on several provider resources or protocols.

\noindent\textbf{\textit{P6: Machine-readable and versioned specification.}}
OpenAPI~3.1 and JSON Schema formally define the outward interface. This
supports validation, generated clients, interactive documentation, conformance
testing, and explicit version management.

\noindent\textbf{\textit{P7: Time-specific characterization.}}
Time-dependent device observations include provider identifiers and observation
or validity timestamps where available. Their association with an execution is
qualified as provider-linked, configuration-linked, timestamp-matched,
nearest-available, or current-at-retrieval.

\noindent\textbf{\textit{P8: Explicit evidence and availability.}}
Each evaluated value can be classified as provider-supplied,
aggregator-normalized, provider-pass-through, derived, application-captured,
unavailable, not applicable, or not verifiable. Derived values identify their
inputs and derivation rule.

\noindent\textbf{\textit{P9: Execution, result, and retention transparency.}}
The model records job lifecycle information, result representations,
transformations, storage references, and retention state. It distinguishes
information that was never collected from information that was later
transformed, archived, or deleted.

\noindent\textbf{\textit{P10: Software and access context.}}
The record identifies relevant client, framework, compiler, adapter, API or
protocol versions, access paths, and retrieval times. Information not retained
by a provider may be supplied by the submitting or retrieving application.

\noindent\textbf{\textit{P11: Explicit evolution and backward compatibility.}}
API and schema versions are externally visible. Incompatible changes require a
new supported version, while additive changes follow documented compatibility
and deprecation policies.

\subsection{Core API specification}
\label{subsec:apispec}

The conceptual API is primarily read-oriented. Provider adapters reconstruct
the best-available provenance record from upstream provider resources.
Application-side capture operations allow source programs, compilation
artifacts, software environments, and other locally available context to be
associated with an external provider job.

The API assigns canonical identifiers to records, jobs, devices, artifacts,
and characterization snapshots while retaining provider-native identifiers.
It also distinguishes the platform through which an execution was managed from
the provider operating the physical hardware. A single outward-facing request
may require several upstream adapter operations; the API centralizes this
aggregation rather than implying that all provenance originates from one
provider endpoint.

Table~\ref{tab:apiendpoints} summarizes the principal endpoints in the
conceptual API. The evaluated reference prototype implements the two operations
needed for provenance-record assembly and retrieval; the remaining endpoints
define the broader intended interface.

\begin{table}[!htbp]
\centering
\caption{Core endpoints in the unified quantum provenance API}
\label{tab:apiendpoints}
\footnotesize
\setlength{\tabcolsep}{3.5pt}
\renewcommand{\arraystretch}{1.12}

\begin{tabularx}{\textwidth}{
  >{\RaggedRight\arraybackslash\ttfamily}p{1.05cm}
  >{\RaggedRight\arraybackslash}p{6.25cm}
  >{\RaggedRight\arraybackslash}X
}
\toprule
\textbf{\normalfont Method}
& \textbf{Endpoint}
& \textbf{Purpose} \\
\midrule

\multicolumn{3}{@{}l}{\textit{Provider and device discovery}} \\

GET
& \path|/v1/providers|
& List provider platforms and cloud aggregators. \\

GET
& \path|/v1/providers/{provider_id}|
& Retrieve provider identity, available access methods, and adapter
capabilities. \\

GET
& \path|/v1/providers/{provider_id}/devices|
& List devices exposed through a provider platform. \\

GET
& \path|/v1/devices/{device_id}|
& Retrieve normalized device identity, capabilities, status, and supported
program representations. \\

\midrule
\multicolumn{3}{@{}l}{\textit{Device characterization}} \\

GET
& \path|/v1/devices/{device_id}/characterizations|
& List calibration, quality-metric, configuration, certificate, and other
characterization snapshots. \\

GET
& \path|/v1/devices/{device_id}/characterizations/{snapshot_id}|
& Retrieve a specific immutable or provider-identified characterization
snapshot. \\

GET
& \path|/v1/devices/{device_id}/characterizations?at={timestamp}|
& Retrieve the snapshot applicable or temporally closest to a specified time. \\

\midrule
\multicolumn{3}{@{}l}{\textit{Job provenance}} \\

GET
& \path|/v1/jobs/{job_id}|
& Retrieve normalized job identity, target, lifecycle status, and execution
summary. \\

GET
& \path|/v1/jobs/{job_id}/provenance|
& Retrieve the best-available unified provenance record. \\

GET
& \path|/v1/jobs/{job_id}/artifacts|
& List source, submitted, compiled, executed, environment, and result
artifacts associated with the job. \\

GET
& \path|/v1/jobs/{job_id}/results|
& Retrieve normalized result metadata and references to result artifacts. \\

GET
& \path|/v1/jobs/{job_id}/provider-native|
& Retrieve sanitized provider-native metadata or references to retained native
responses. \\

\midrule
\multicolumn{3}{@{}l}{\textit{Application-side capture}} \\

POST
& \path|/v1/provenance-records|
& Create a provenance record for an external provider job and supply
application-captured context. \\

POST
& \path|/v1/provenance-records/{record_id}/artifacts|
& Associate source programs, compilation artifacts, environment manifests, or
other client-captured evidence with an existing record. \\

\bottomrule
\end{tabularx}
\end{table}
\FloatBarrier

\subsubsection{Provenance response structure}

The \texttt{/v1/jobs/\{job\_id\}/provenance} endpoint returns the
best-available normalized record for an execution. The response combines
common provenance fields, artifact lineage, device-characterization
associations, technology-specific extensions, and evidence describing the
origin and availability of individual attributes.

Listing~\ref{lst:provenancesummary} presents the top-level structure. A
complete illustrative response for the conceptual API is provided in the
research artifact~\cite{peltonen2026artifact}. It is not a verbatim
response from one provider. Generated records from the evaluated prototype are
provided separately in the same artifact.

\begin{lstlisting}[
  caption={Top-level unified provenance response structure},
  label={lst:provenancesummary},
  language=json,
  basicstyle=\ttfamily\scriptsize
]
{
  "schema_version": "1.0.0",
  "record_id": "qprov-record-7d4fd53a",
  "generated_at": "2026-07-12T14:05:22Z",
  "job": { ... },
  "device": { ... },
  "program": {
    "representation": "gate_model",
    "summary": { ... },
    "artifacts": [ ... ]
  },
  "compilation": { ... },
  "execution": { ... },
  "characterization": { ... },
  "results": { ... },
  "software": { ... },
  "extensions": { ... },
  "evidence": [ ... ],
  "provider_native": { ... }
}
\end{lstlisting}

The response retains both canonical and external provider identifiers and
distinguishes the execution platform from the hardware operator. For example,
a QuEra execution managed through Amazon Braket identifies Amazon Braket as
the platform and QuEra as the hardware provider.

The \texttt{program.artifacts} collection represents the program and result
lineage. Its principal roles are:

\begin{itemize}
    \item \texttt{source}: the original application-level program;
    \item \texttt{submitted}: the representation sent to the provider;
    \item \texttt{compiled}: a device-targeted representation;
    \item \texttt{executed}: the representation confirmed as executed, when
    available;
    \item \texttt{environment}: a software or dependency manifest; and
    \item \texttt{result}: a raw or transformed result artifact.
\end{itemize}

This prevents a provider-retained QIR, OpenQASM~\cite{cross2022openqasm3}, Quil~\cite{smith2017practicalquantuminstructionset}, pulse, analog, or native
representation from being incorrectly presented as the application's original
source program. Logical-to-physical mappings use structured entries so that
the representation can also accommodate atom-to-trap mappings, photonic modes,
annealing embeddings, and mappings produced at several compilation stages.

The \texttt{evidence} collection describes the origin and availability of
normalized values:

\begin{itemize}
    \item \texttt{provider\_supplied}: obtained directly from a hardware
    provider;
    \item \texttt{aggregator\_normalized}: normalized by a cloud aggregator;
    \item \texttt{provider\_passthrough}: relayed by an aggregator without
    semantic normalization;
    \item \texttt{derived}: calculated from other values;
    \item \texttt{application\_captured}: supplied by the local application or
    compiler;
    \item \texttt{unavailable}: applicable but not exposed or captured;
    \item \texttt{not\_applicable}: not meaningful for the technology or
    workflow; and
    \item \texttt{not\_verifiable}: documented or indicated but not verifiable
    from the available evidence.
\end{itemize}

Derived values identify their source paths and derivation method. Unavailable
values include a reason, such as
\texttt{not\_exposed\_by\_provider},
\texttt{not\_captured\_by\_application},
\texttt{discarded\_by\_retention\_policy}, or
\texttt{insufficient\_source\_data}. The model can therefore represent why an
attribute is absent rather than relying only on omitted or null fields.

Provider-native information is stored separately from the normalized record or
referenced through a protected artifact location. This preserves information
not yet represented in the common schema while allowing credentials, signed
locations, personal identifiers, and other sensitive fields to be redacted.

The result model likewise preserves differences in representation. Observed
measurement counts, provider-returned probabilities, raw shots,
quasi-probabilities, and transformed results are not treated as
interchangeable. If a derived representation is provided, the original
representation, conversion method, and evidence status remain explicit.

The endpoint thus provides one application-facing record without implying that
all values have the same origin, completeness, or evidential strength. It
cannot create information absent from both provider interfaces and
application-side capture; such limitations are represented explicitly.

\subsubsection{Device characterization model}

The
\texttt{/v1/devices/\{device\_id\}/characterizations/\{snapshot\_id\}}
endpoint returns a timestamped device-characterization snapshot.
\emph{Characterization} is used as an umbrella term for calibration records,
processor configurations, quality-metric sets, device certificates, solver
properties, and other time-dependent observations.

Stable capabilities, such as nominal capacity, supported program formats,
native operations, topology constraints, and programmable ranges, belong to
the device resource. Time-dependent measurements belong to characterization
snapshots. Job-specific values, such as shots, annealing time, chain strength,
optimization level, or selected pulse parameters, remain in the program,
compilation, or execution sections.

A characterization snapshot contains:

\begin{itemize}
    \item canonical and provider-native snapshot identifiers;
    \item a snapshot kind, such as \texttt{calibration},
    \texttt{quality\_metrics}, \texttt{processor\_configuration},
    \texttt{device\_certificate}, or \texttt{solver\_properties};
    \item observation, validity, and retrieval timestamps;
    \item the association between the snapshot and the execution;
    \item typed metrics with values, units, targets, uncertainty, timestamps,
    and evidence; and
    \item namespaced technology-specific characterization.
\end{itemize}

A direct provider association is classified as
\texttt{provider\_linked} or \texttt{configuration\_linked}. Where no direct
link exists, an adapter may select a timestamp-matched or nearest-available
snapshot and record its temporal difference from the execution. A snapshot
retrieved after execution without historical validity is classified as
\texttt{current\_at\_retrieval} and must not be presented as the state used
during the execution.

For gate-based systems, common metrics include coherence times, operation
errors or fidelities, operation durations, readout errors, and
state-preparation and measurement errors. Metrics remain distinct unless the
provider defines an explicit mathematical relationship between them.
Listing~\ref{lst:characterizationmetric} shows an example.

\begin{lstlisting}[
  caption={Example normalized device-characterization metric},
  label={lst:characterizationmetric},
  language=json,
  basicstyle=\ttfamily\scriptsize
]
{
  "metric": "gate_error",
  "operation": "cx",
  "targets": [45, 46],
  "value": 0.0023,
  "unit": "dimensionless",
  "uncertainty": null,
  "observed_at": "2026-07-17T09:00:00Z",
  "evidence": {
    "status": "provider_supplied",
    "provider_field": "gate_error"
  }
}
\end{lstlisting}

Technology-specific characterization may include:

\begin{itemize}
    \item annealer working graphs, inactive components, and flux-bias or
    readout characteristics;
    \item photonic source efficiency, loss, indistinguishability, multiphoton
    emission, and detector characteristics;
    \item neutral-atom preparation, readout, trap-layout, interaction, and
    pulse-control characteristics; and
    \item gate-based pulse calibrations, native-operation parameters,
    crosstalk metrics, and processor-configuration identifiers.
\end{itemize}

Technology-specific job settings remain outside the characterization snapshot.
For example, an annealer's supported schedule range is a device capability,
the selected schedule and chain strength are execution or compilation
settings, and the working graph observed at a particular time is
characterization.

Together, the provenance and characterization models define interoperable
common resources while retaining technology-specific semantics. Normalization
therefore makes heterogeneous records accessible without implying that every
provider exposes the same data.

\subsubsection{Security and privacy considerations}
Provenance records aggregate information that may be sensitive: submitted
programs can embed proprietary algorithms, provider-native responses may
contain account identifiers, signed storage locations, or billing data, and
software-environment manifests can reveal internal infrastructure. The
proposed design addresses this at three levels. First, provider-native evidence is
stored separately from the normalized record and passes through a redaction layer 
that masks credential-bearing fields before a record is emitted; the reference 
prototype implements and tests this behavior. Second, signed storage locations and account identifiers are additionally kept out of emitted records by referencing provider-native material through locators and content hashes rather than inlining payloads. Third, the adoption strategy assigns
authentication, authorization, encryption, and retention policies to
production deployments and their governance (Section~\ref{subsec:adoption}).
A full threat model for multi-tenant provenance services, including access
control for provider-native evidence and secure sharing of records between
organizations, remains future work.

\subsection{Provider adapter pattern}
\label{subsec:adapters}

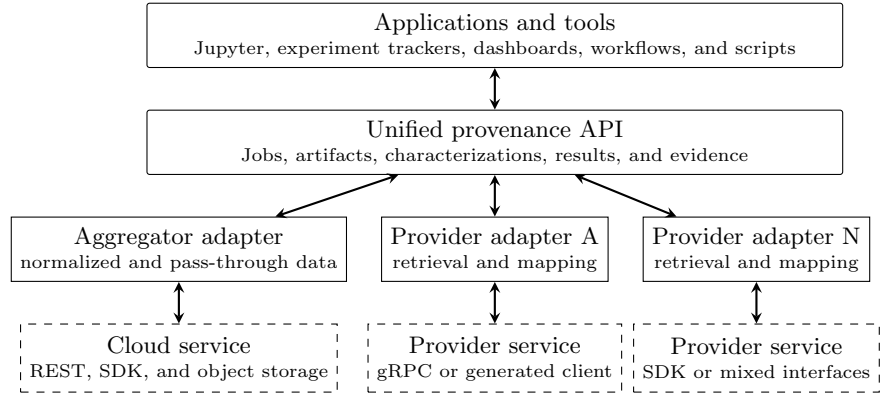
\begin{figure}[!htbp]
\centering
\begin{tikzpicture}[
    node distance=0.55cm and 0.45cm,
    layer/.style={
        rectangle,
        draw,
        rounded corners=1pt,
        minimum height=0.85cm,
        align=center,
        font=\scriptsize
    },
    adapter/.style={
        rectangle,
        draw,
        minimum width=2.45cm,
        minimum height=0.85cm,
        align=center,
        font=\scriptsize
    },
    upstream/.style={
        rectangle,
        draw,
        dashed,
        minimum width=2.45cm,
        minimum height=0.9cm,
        align=center,
        font=\scriptsize
    },
    arrow/.style={<->, >=stealth, thick}
]

\node[layer, minimum width=9.2cm] (apps)
{Applications and tools\\
\footnotesize Jupyter, experiment trackers, dashboards, workflows, and scripts};

\node[layer, minimum width=9.2cm, below=of apps] (api)
{Unified provenance API\\
\footnotesize Jobs, artifacts, characterizations, results, and evidence};

\node[adapter, below=of api] (ibm)
{Provider adapter A\\
\footnotesize retrieval and mapping};

\node[adapter, left=of ibm] (aggregator)
{Aggregator adapter\\
\footnotesize normalized and pass-through data};

\node[adapter, right=of ibm] (providerN)
{Provider adapter N\\
\footnotesize retrieval and mapping};

\node[upstream, below=of aggregator] (upstreamA)
{Cloud service\\
\footnotesize REST, SDK, and object storage};

\node[upstream, below=of ibm] (upstreamB)
{Provider service\\
\footnotesize gRPC or generated client};

\node[upstream, below=of providerN] (upstreamC)
{Provider service\\
\footnotesize SDK or mixed interfaces};

\draw[arrow] (apps) -- (api);
\draw[arrow] (api) -- (aggregator);
\draw[arrow] (api) -- (ibm);
\draw[arrow] (api) -- (providerN);
\draw[arrow] (aggregator) -- (upstreamA);
\draw[arrow] (ibm) -- (upstreamB);
\draw[arrow] (providerN) -- (upstreamC);

\end{tikzpicture}
\caption{Provider-adapter architecture for exposing heterogeneous upstream
interfaces through the unified provenance API.}
\label{fig:adapter}
\end{figure}

Immediate native adoption of one provenance interface by every provider is
unlikely. The proposed architecture therefore uses provider adapters to map
existing services into the common API. As shown in
Figure~\ref{fig:adapter}, applications communicate only with the unified
interface, while adapters retrieve and qualify the required upstream
resources.

An adapter has six principal responsibilities:

\begin{enumerate}
    \item \textbf{Retrieve and aggregate resources.}
    The adapter obtains available job, program, device, characterization,
    result, and provider-native information. These resources may originate from
    several endpoints, SDK calls, generated clients, or object-storage
    locations.

    \item \textbf{Normalize semantics.}
    Provider terminology, identifiers, timestamps, structures, and units are
    mapped into the common schema without equating metrics whose meanings
    differ. Transformations retain their original values and assumptions where
    needed.

    \item \textbf{Preserve identity and lineage.}
    Canonical identifiers are assigned while provider-native and
    aggregator-generated identifiers remain available. Program artifacts are
    related without assuming that a provider-retained representation is the
    original application source.

    \item \textbf{Associate characterization.}
    The adapter prefers characterization explicitly linked by the provider. If
    none exists, it may select a temporally matched or nearest-available
    snapshot and report the association method and time difference.

    \item \textbf{Report evidence and gaps.}
    Values are classified by origin and availability. Missing data, failed
    retrievals, application-captured context, derived fields, and inapplicable
    concepts are represented explicitly.

    \item \textbf{Preserve provider-native information safely.}
    Native data may be retained or referenced when permitted, after redacting
    credentials, signed storage locations, personal identifiers, and other
    sensitive fields.
\end{enumerate}

A provenance request may require several upstream operations. Failure to
retrieve one optional resource, such as historical characterization, must not
prevent other sections from being returned. Adapters therefore support partial
success and report which operations failed or which sections may be
incomplete. Each retrieved section records its retrieval time. Resource-specific caching
can be used where appropriate: stable device capabilities may be cached longer
than job status or characterization information, while immutable snapshots can
be reused once retained.

Technology-specific data are placed in namespaced extension objects. Their
location follows their semantics: an annealer working graph belongs to device
characterization, embedding belongs to compilation, and the selected annealing
schedule belongs to execution. This avoids forcing technology-specific
concepts into unrelated gate-model fields.

Adapters declare their supported capabilities, such as whether they can
retrieve submitted programs, historical characterization, raw results,
provider-native records, or execution-linked calibration. The existence of an
adapter therefore does not imply complete provenance coverage. A common conformance 
suite validates schema structure, identifiers, units, evidence states, redaction, 
partial-record behavior, and extension semantics. Adapter and mapping versions are 
included in generated records so that normalization behavior can be reproduced.

Adapters may be maintained by providers, platform operators, or open-source
communities. Regardless of ownership, reliable maintenance requires versioned
releases, documented responsibility, conformance tests, compatibility
declarations, and monitoring of upstream interface changes. The controlled
change experiment in Section~\ref{sec:casestudy} evaluates this adapter
boundary in the reference prototype.

\subsection{Adoption and evolution strategy}
\label{subsec:adoption}

The design supports incremental adoption rather than requiring coordinated
provider migration. The specification, adapters, and optional native
implementations can coexist.

\noindent\textbf{\textit{Technical baseline.}}
This work establishes an initial technical baseline consisting of a conceptual
API design, an OpenAPI~3.1 contract for the evaluated prototype, provider
adapters representing aggregator-mediated and direct-provider access,
canonical examples, a versioned attribute inventory, and automated
conformance and drift checks. The reference prototype, implemented within
QMill's provider-executor architecture and evaluated in
Section~\ref{sec:casestudy}, demonstrates that existing provider interfaces
can be mapped into a common evidence-aware record without modifying the
provider execution path.

The evaluated implementation is intentionally narrower than the complete
conceptual API. It provides in-process provenance-record assembly and retrieval
rather than a production HTTP deployment, durable storage, full provider and
device discovery, or native provider adoption.

\noindent\textbf{\textit{Community review and governed evolution.}}
The next stage is broader review by hardware providers, cloud aggregators,
application developers, researchers, and relevant standards communities.
Governance should define:

\begin{itemize}
    \item ownership and versioning of the common schema;
    \item registration and evolution of technology-specific extensions;
    \item security and privacy requirements for provider-native evidence;
    \item compatibility and deprecation policies;
    \item controlled vocabularies for metrics, units, and identifiers; and
    \item conformance profiles for different implementation capabilities.
\end{itemize}

Conformance profiles allow an implementation to support, for example, job
provenance without claiming historical characterization or complete artifact
lineage. Applications can inspect these capabilities before requesting a
record.

\noindent\textbf{\textit{Production integration and optional native adoption.}}
Production implementations would add authentication, authorization, durable
storage, encryption, artifact-retention policies, caching, rate-limit handling,
monitoring, and refresh semantics. Workflow systems and local compilers would
supply application-captured source, compilation, and software context.

Providers and aggregators could subsequently expose the common schema natively
or maintain official adapters. Native adoption may improve freshness and
reduce adapter complexity, especially for execution-linked characterization
and provider-side compilation. It is not a prerequisite for initial use:
maintained adapters can continue to provide a stable interface over existing
services.

Providers may also have competitive reasons to limit exposure of detailed 
characterization or compilation data. The proposed model accommodates this: 
conformance profiles and explicit unavailability semantics allow a provider 
to participate at a chosen disclosure level without breaking the common contract, 
rather than making full disclosure a precondition for interoperability.

\noindent\textbf{\textit{Long-term ecosystem use.}}
With broader adoption, evidence-aware records could support reproducibility
checks, experiment tracking, calibration-aware device selection, temporal
performance analysis, automated execution comparison, and provenance-quality
assessment. Such uses must continue to account for evidence source,
availability, freshness, and temporal association rather than assuming that
all normalized values are equally complete.

This strategy lowers the initial coordination barrier: the common contract and
adapters provide immediate integration value, conformance profiles make
partial implementations explicit, and governed evolution allows providers and
technologies to adopt or extend the model incrementally.

\section{Reference prototype and demonstration at QMill}
\label{sec:casestudy}

To demonstrate the practical relevance of unified quantum provenance access,
we examine its application in the software architecture of
QMill\footnote{\url{https://qmill.com/en}}, a Finnish company developing
quantum algorithms intended to provide utility during the NISQ era. QMill
evaluates algorithms across multiple quantum computers and provider platforms.
Recording the program, compilation, execution, device-characterization, result,
and software context of each run is therefore important for comparing
executions across devices and time periods and for retaining evidence of the
conditions under which a result was produced.

The coexistence of multiple quantum-computing technologies~\cite{gill}
requires software infrastructure capable of adapting to technological
diversity and uncertainty. Consequently, supporting a provider platform involves both execution
integration and provenance integration. Execution integration translates and
submits programs, monitors jobs, and retrieves results. Provenance integration
collects and normalizes the contextual information required to interpret,
compare, and reproduce those executions. The unified API proposed in this work
primarily addresses the latter responsibility; it does not replace the
provider-specific interfaces used to execute quantum programs.

QMill uses a provider-executor pattern to isolate platform-specific execution
logic. Nevertheless, provenance retrieval must still account for differences
in program representations, job states, device descriptions,
characterization data, timestamps, result formats, and software dependencies.
The reference prototype extends the existing pattern with a separate
provider-independent provenance service. Provider-specific execution remains
unchanged, whereas provenance retrieval and normalization are placed behind a
common interface.

\subsection{Prototype scope}
\label{subsec:prototype-scope}

The principal demonstration cases represent two structurally different
provider-integration styles. The first uses Amazon Braket, an AWS-managed
aggregation platform through which devices from several hardware providers can
be accessed using a common task and device model. The second uses IBM Quantum
as a direct provider platform through the Qiskit IBM Runtime interface. These
cases were selected to maximize architectural contrast while keeping the
prototype focused.

The Amazon Braket case represents a task routed to an IonQ device. Its
provenance record therefore distinguishes Amazon Braket as the platform through
which the execution was managed from IonQ as the operator of the underlying
hardware. The IBM case instead represents IBM as both the provider platform
and the hardware operator. A direct IonQ mapping was additionally implemented
as a supplementary conformance case. It was used to test result semantics that
differ from both the Braket and IBM cases, particularly the distinction between
provider-returned probabilities and observed measurement counts.

Table~\ref{tab:prototypecases} summarizes the implemented cases.

\begin{table}[!htbp]
\centering
\caption{Provider-integration models included in the reference prototype.}
\label{tab:prototypecases}
\footnotesize
\setlength{\tabcolsep}{4pt}
\renewcommand{\arraystretch}{1.1}

\begin{tabularx}{\textwidth}{p{2.4cm} p{2.0cm} p{2.7cm} X}
\toprule
\textbf{Platform}
& \textbf{Hardware provider}
& \textbf{Integration model}
& \textbf{Provenance characteristics} \\
\midrule

Amazon Braket
& IonQ
& Cloud aggregator
& Common task and device models, underlying-provider attribution,
provider-pass-through artifacts, current-at-retrieval characterization, and
application-captured compilation context. \\

IBM Quantum
& IBM
& Direct provider platform
& Provider-specific job, backend, result, and characterization models,
combined with application-captured Qiskit compilation and software context. \\

IonQ Quantum Cloud
& IonQ
& Direct provider platform
& Provider-supplied probability histogram, derived count estimates,
provider-linked characterization, and provider-supplied execution duration. \\

\bottomrule
\end{tabularx}
\end{table}

The prototype is limited to provenance retrieval for representative completed
jobs. It does not implement unified job submission, comprehensive device
discovery, historical characterization browsing, billing, production
authentication infrastructure, or every endpoint and field in the complete
conceptual API. The principal research objectives are to evaluate common-schema
feasibility, explicit evidence attribution, partial-record behavior, adapter
isolation, and application-facing integration simplification.

The evaluation is entirely offline and uses deterministic, sanitized fixtures
constructed from publicly available SDK documentation, public data models,
source code, schemas, and examples. No provider credentials or network access
are required. The prototype therefore evaluates the structural feasibility of
normalization rather than the runtime completeness, retention behavior, or
reliability of authenticated provider services.

\subsection{Reference implementation}
\label{subsec:prototype-implementation}

QProv's implementation\footnote{\url{https://github.com/UST-QuAntiL/qprov}}
is a capture-and-store provenance system centered on its own data model and
collection agents, whereas this work targets a provider-facing
retrieval/normalization contract integrated into an existing industrial provider-
executor architecture; the schema-level mapping (Table~\ref{tab:qprovmapping}) keeps 
the two interoperable, and integrating the unified API as a data source for a QProv 
deployment is plausible future work.

The reference implementation extends QMill's existing provider-executor
architecture with a sibling provenance-adapter abstraction, an adapter
registry, and an in-process provenance service. The normative contract is a
reduced OpenAPI~3.1 specification implementing two operations:

\begin{center}
\begin{tabular}{ll}
\texttt{POST} &
\path|/v1/provenance-records| \\[2pt]
\texttt{GET} &
\path|/v1/jobs/{job_id}/provenance|
\end{tabular}
\end{center}

The first operation registers an external provider job and assembles its
provenance record. Its request contains a provider identifier, an external job
identifier, optional provider-routing information, and optional
application-captured context. The second operation retrieves an assembled
record using its canonical identifier. The prototype exposes equivalent
in-process service methods; it does not include a separate HTTP server or
persistent database.

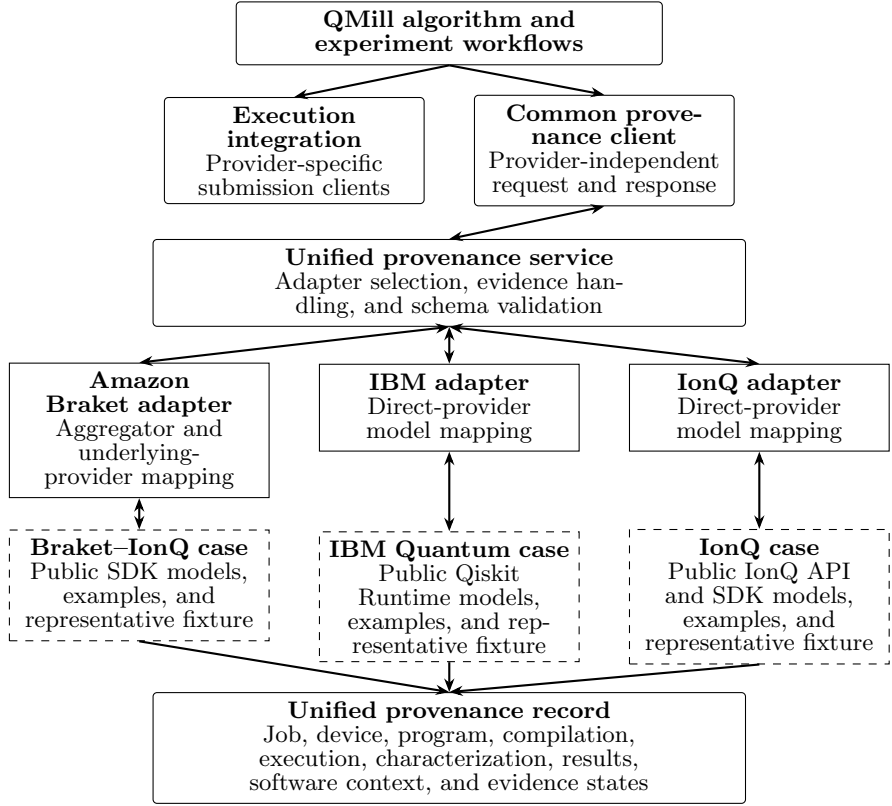
\begin{figure}[!htbp]
\centering

\begin{tikzpicture}[
    workflow/.style={
        rectangle,
        draw,
        rounded corners=1.5pt,
        text width=5.4cm,
        minimum height=0.75cm,
        align=center,
        font=\scriptsize
    },
    qmill/.style={
        rectangle,
        draw,
        rounded corners=1.5pt,
        text width=3.2cm,
        minimum height=1.0cm,
        align=center,
        font=\scriptsize
    },
    common/.style={
        rectangle,
        draw,
        rounded corners=1.5pt,
        text width=7.6cm,
        minimum height=0.85cm,
        align=center,
        font=\scriptsize
    },
    adapter/.style={
        rectangle,
        draw,
        text width=3.2cm,
        minimum height=1.0cm,
        align=center,
        font=\scriptsize
    },
    source/.style={
        rectangle,
        draw,
        dashed,
        text width=3.2cm,
        minimum height=1.0cm,
        align=center,
        font=\scriptsize
    },
    flow/.style={
        -{Stealth[length=1.8mm]},
        thick
    },
    exchange/.style={
        {Stealth[length=1.8mm]}-{Stealth[length=1.8mm]},
        thick
    }
]

\node[workflow] (workflow) at (0,0)
{\textbf{QMill algorithm and experiment workflows}};

\node[qmill] (execution) at (-2.05,-1.55)
{\textbf{Execution integration}\\
Provider-specific submission clients};

\node[qmill] (client) at (2.05,-1.55)
{\textbf{Common provenance client}\\
Provider-independent request and response};

\node[common] (service) at (0,-3.30)
{\textbf{Unified provenance service}\\
Adapter selection, evidence handling, and schema validation};

\node[adapter] (braket) at (-4.10,-5.25)
{\textbf{Amazon Braket adapter}\\
Aggregator and underlying-provider mapping};

\node[adapter] (ibm) at (0,-4.95)
{\textbf{IBM adapter}\\
Direct-provider model mapping};

\node[adapter] (ionq) at (4.10,-4.95)
{\textbf{IonQ adapter}\\
Direct-provider model mapping};

\node[source] (braketsource) at (-4.10,-7.30)
{\textbf{Braket--IonQ case}\\
Public SDK models, examples, and representative fixture};

\node[source] (ibmsource) at (0,-7.45)
{\textbf{IBM Quantum case}\\
Public Qiskit Runtime models, examples, and representative fixture};

\node[source] (ionqsource) at (4.10,-7.45)
{\textbf{IonQ case}\\
Public IonQ API and SDK models, examples, and representative fixture};

\node[common] (record) at (0,-9.45)
{\textbf{Unified provenance record}\\
Job, device, program, compilation, execution, characterization,
results, software context, and evidence states};

\draw[flow]
    (workflow.south)
    -- (execution.north);

\draw[flow]
    (workflow.south)
    -- (client.north);

\draw[exchange]
    (client.south)
    -- (service.north);

\draw[exchange]
    (service.south)
    -- (braket.north);

\draw[exchange]
    (service.south)
    -- (ibm.north);

\draw[exchange]
    (service.south)
    -- (ionq.north);    

\draw[exchange]
    (braket.south)
    -- (braketsource.north);

\draw[exchange]
    (ibm.south)
    -- (ibmsource.north);

\draw[exchange]
    (ionq.south)
    -- (ionqsource.north);

\draw[flow]
    (braketsource.south)
    -- (record.north);

\draw[flow]
    (ibmsource.south)
    -- (record.north);

\draw[flow]
    (ionqsource.south)
    -- (record.north);
    
\end{tikzpicture}

\caption{Reference-prototype architecture used in the QMill demonstration.
QMill accesses provenance through one common client, while separate Amazon Braket, IBM, 
and IonQ adapters normalize representative public provider structures
into the unified provenance record. The execution path remains
provider-specific and is outside the prototype's implementation scope.}
\label{fig:qmillprototype}
\end{figure}

Figure~\ref{fig:qmillprototype} shows the implementation architecture.

The implementation contains four principal parts:

\begin{enumerate}
    \item an OpenAPI contract and corresponding canonical domain model;

    \item a provider-independent provenance service responsible for canonical
    identifiers, application-context merging, record assembly, and operation
    accounting;

    \item a registry that selects provider-specific provenance adapters without
    modifying the existing execution path; and

    \item fixture-backed Braket, IBM, and IonQ adapters that map heterogeneous
    provider structures into the common record.
\end{enumerate}

The implemented record covers canonical and provider-native job identity,
provider-platform and hardware-provider identity, target-device information,
program artifacts, compilation context, execution timing, characterization,
results, software context, evidence qualifiers, warnings, and logical
retrieval operations. The OpenAPI schemas use
\texttt{additionalProperties: false} to detect undocumented changes in the
implemented contract.

The record supports the evidence states
\texttt{provider\_supplied},
\texttt{aggregator\_normalized},
\texttt{provider\_passthrough},
\texttt{derived},
\texttt{application\_captured},
\texttt{unavailable},
\texttt{not\_applicable}, and
\texttt{not\_verifiable}. The evaluation additionally uses
\texttt{unclassified} to identify an emitted attribute that lacks the expected
evidence qualifier. No completed demonstration record contained an
unclassified inventory attribute.

Application-captured context includes the source-program hash and reference,
compiler or transpiler identity, optimization level, logical-to-physical
mapping, provider-client version, Python version, and framework versions. The
prototype supplies deterministic application context for the Braket and IBM
cases. These values are marked as application-captured and are not presented as
provider-returned evidence.

The prototype also distinguishes the provider platform from the hardware
operator. In the Braket case, the canonical device record contains:

\begin{center}
\texttt{platform\_provider = amazon\_braket}
\qquad
\texttt{hardware\_provider = ionq}.
\end{center}

For the direct IBM and IonQ cases, the platform and hardware operator refer to
the same provider.

Result representations retain their original semantics. Braket and IBM provide
integer measurement counts, which are stored in
\texttt{measurement\_counts}. The direct IonQ fixture instead provides a
probability histogram. These probabilities are retained in
\texttt{measurement\_probabilities}, with integer state identifiers rewritten
as fixed-width bit strings. Optional
\texttt{estimated\_measurement\_counts} are derived independently as
\(\operatorname{round}(p_i N)\), where \(p_i\) is a provider-supplied outcome
probability and \(N\) is the requested shot count. These estimates are
explicitly distinguished from observed counts and are not adjusted to force
their sum to equal the requested shot count.

Characterization association is also represented explicitly. The Braket
fixture provides a device snapshot classified as
\texttt{current\_at\_retrieval}. The IBM record contains the nearest available
backend-properties snapshot and reports its temporal difference from the job
time without claiming that the snapshot was in force during execution. The
IonQ record contains a characterization associated through a provider-defined
identifier and is therefore classified as \texttt{provider\_linked}.

\subsection{Evaluation procedure}
\label{subsec:prototype-evaluation}

The evaluation uses a fixed, versioned inventory of 21 provenance attributes.
The inventory covers job identity and timing, device identity, platform and
hardware attribution, compiled artifacts, execution durations,
characterization, result representations, compilation context, and software
context. Every inventory attribute is assigned exactly one classification for
each provider case.

The evaluation applies four complementary procedures.

\noindent\textbf{\textit{Schema and normalization feasibility.}}
The completed Braket, IBM, and IonQ records are validated against the same
OpenAPI contract. The fixed inventory is used to count provider-supplied,
pass-through, derived, application-captured, unavailable, and inapplicable
attributes. Raw evidence-entry counts are reported separately because an
evidence object is not equivalent to one inventory attribute.

\noindent\textbf{\textit{Partial-record behavior.}}
For each provider, a second fixture omits characterization data. The experiment
tests whether the remaining job, device, execution, result, and software
information can still be returned as a schema-valid record while
characterization is explicitly classified as unavailable.

\noindent\textbf{\textit{Provider-input change isolation.}}
An IBM-specific input field,
\texttt{device.provider\_properties.num\_qubits}, is renamed to
\texttt{qubit\_total}. The experiment verifies that the resulting IBM record
remains schema-valid, the affected canonical attribute degrades to unavailable,
and the Braket and IonQ records remain unchanged. Supporting the renamed field
requires a localized change in the IBM adapter rather than a change to the
shared contract or other provider adapters.

\noindent\textbf{\textit{Schema evolution.}}
An optional \texttt{cost} attribute is added to an in-memory copy of the
contract. Existing records must remain valid under the extended schema, while
a record containing the new attribute must be rejected by the original closed
schema and accepted by the extended schema. This evaluates additive
contract evolution separately from provider-input change isolation.

The implementation also reports handwritten source lines and logical operation
counts. Blank lines and comment-only lines are excluded. Generated examples,
fixtures, tests, and documentation are not included in the implementation
figures. The operation counts describe logical fixture-backed calls rather than
measured network requests.

\subsection{Demonstration results}
\label{subsec:prototype-results}

All three completed-job records validated against the same reduced OpenAPI
contract. Table~\ref{tab:prototypecoverage} reports the classifications of the
21 fixed inventory attributes.

\begin{table}[!htbp]
\centering
\caption{Classification of the fixed provenance-attribute inventory.}
\label{tab:prototypecoverage}
\footnotesize
\setlength{\tabcolsep}{3.2pt}
\renewcommand{\arraystretch}{1.08}

\begin{tabularx}{\textwidth}{
    l
    >{\centering\arraybackslash\hsize=0.85\hsize}X
    >{\centering\arraybackslash\hsize=0.95\hsize}X
    >{\centering\arraybackslash\hsize=0.85\hsize}X
    >{\centering\arraybackslash\hsize=1.2\hsize}X
    >{\centering\arraybackslash\hsize=1.2\hsize}X
    >{\centering\arraybackslash\hsize=0.75\hsize}X
    >{\centering\arraybackslash\hsize=1.2\hsize}X
}
\toprule
\textbf{Case}
& \textbf{Provider}
& \textbf{Pass-through}
& \textbf{Derived}
& \textbf{Application}
& \textbf{Unavailable}
& \textbf{N/A}
& \textbf{Unclassified} \\
\midrule
Braket-IonQ  & 10 & 1 & 2 & 4 & 2 & 2 & 0 \\
IBM Quantum   & 11 & 1 & 3 & 4 & 0 & 2 & 0 \\
Direct IonQ   & 12 & 1 & 2 & 0 & 5 & 1 & 0 \\
\bottomrule
\end{tabularx}
\end{table}

The inventory results show that the common contract accommodates materially
different provenance profiles without concealing their differences. The
Braket-IonQ case distinguishes the aggregation platform from the hardware
operator and combines provider data with application-captured compilation and
software context. The IBM case demonstrates direct-provider normalization and
derives execution durations from provider-supplied timestamps. The direct-IonQ
case preserves provider-returned probabilities and represents estimated counts
separately as derived values. Attributes absent because they do not apply to a
provider representation are classified as not applicable rather than
unavailable. No completed case contained an unclassified inventory attribute.

Table~\ref{tab:prototypeimplementationresults} reports implementation size and
logical operations. The shared implementation comprises the canonical model,
adapter abstraction, registry, service, redaction utilities, and contract
validation. Lines of code (LOC) are reported as an indicator of where implementation responsibility resides, not as a measure of integration effort; adapter sizes are similar across the three structurally different provider integrations, and provider-specific code remains bounded relative to the shared implementation.

\begin{table}[!htbp]
\centering
\caption{Reference-prototype implementation and logical-operation measurements.}
\label{tab:prototypeimplementationresults}
\footnotesize
\setlength{\tabcolsep}{4pt}
\renewcommand{\arraystretch}{1.08}

\begin{tabularx}{\textwidth}{X r r r}
\toprule
\textbf{Component or case}
& \textbf{Handwritten LOC}
& \textbf{QMill-facing operations}
& \textbf{Adapter retrievals} \\
\midrule

Shared provider-independent implementation
& 861 & - & - \\

Amazon Braket adapter
& 428 & 1 & 4 \\

IBM adapter
& 486 & 1 & 4 \\

IonQ adapter
& 526 & 1 & 4 \\

\bottomrule
\end{tabularx}
\end{table}

The measurements separate the shared provider-independent implementation from
the localized adapter code. In every case, QMill performs one logical
provenance-record operation, while the adapter coordinates the upstream
retrieval steps needed to construct the record. Because the evaluation uses
fixture replay, the adapter-operation values represent logical retrievals
rather than measured provider-network requests.

The partial-record experiment succeeded for all three providers. Each record
remained schema-valid when characterization was omitted, and the missing
section was classified as unavailable rather than replaced with an inferred or
current snapshot. No applicable emitted inventory attribute became
unclassified.

The provider-input experiment also satisfied the isolation criterion. After an
IBM-specific qubit-count field was renamed, the IBM record remained valid and
the affected normalized attribute degraded to unavailable. The Braket and IonQ
records remained unchanged, and support for the renamed field was localized to
the IBM adapter.

The schema-evolution experiment showed that all existing example records
remained valid after an optional property was added to an extended contract. A
record using that property was rejected by the original closed schema and
accepted by the extended schema. Strict validation can therefore coexist with
explicitly designed additive evolution.

The implementation and evaluation materials are provided in the accompanying
research artifact~\cite{peltonen2026artifact}. The artifact contains the
source code, reduced OpenAPI contract, sanitized fixtures, fixed attribute
inventory, generated example records, and machine-readable evaluation report.
Automated drift checks verify that the committed examples and report remain
synchronized with the implementation. The complete offline
quantum-provider test suite contained 34 passing tests at the evaluated
revision.

\noindent\textbf{\textit{Result synthesis.}}
The demonstration shows that the reduced contract can represent the three
fixture-backed cases through one QMill-facing workflow while preserving
differences in provider role, result representation, characterization
association, and evidence origin. The controlled experiments further show
that incomplete data and provider-specific input changes can be contained
without invalidating the common record or changing unrelated adapters. The
broader implications are discussed in Section~\ref{sec:discussion}.

\subsection{Prototype limitations}
\label{subsec:prototype-limitations}

The prototype implements a focused subset of the complete conceptual API and
uses deterministic representative fixtures rather than authenticated provider
responses. It therefore evaluates schema and adapter feasibility, not the
runtime completeness of individual provider deployments. Account-specific
permissions, job-data retention, historical characterization availability,
network behavior, rate limits, and region-specific service differences were not
evaluated.

The principal architectural comparison is between Amazon Braket and IBM
Quantum. The direct IonQ adapter is a supplementary semantic and conformance
case rather than a third production integration. Additional direct-provider, 
aggregator, neutral-atom, photonic, and quantum-annealing cases would be required 
to evaluate the generality of the architecture across the complete provider landscape.

The in-process service uses an in-memory record store and does not implement
production authorization, persistence, encryption, refresh policies, or
provider-native artifact storage. Canonical identifier policy, access control
for provider-native evidence, artifact retention, and durable refresh semantics
remain design decisions for a production deployment.

The fixed inventory contains 21 attributes selected for the focused
demonstration. It does not represent every attribute in the complete conceptual
schema. The reported coverage therefore characterizes the implemented prototype
subset rather than the total provenance available from each provider.

Despite these limitations, the demonstration provides concrete and
reproducible evidence that heterogeneous provider structures can be normalized
through one evidence-aware provenance contract, that provider and application
evidence can be combined without conflating their origins, and that
provider-specific changes and missing data can be isolated from QMill's common
provenance workflow.

\section{Evaluation and discussion}
\label{sec:discussion}

This section evaluates the proposed provenance model and unified API against
the objectives defined in Section~\ref{sec:methodology}, answers the research
questions, and discusses the implications and limitations of the results. The
evaluation combines three forms of evidence: the comparative analysis of
public provider interfaces, iterative refinement of the design artifact, and
the reference-prototype demonstration at QMill described in
Section~\ref{sec:casestudy}.

\subsection{Design refinement across iterations}
\label{subsec:design-refinement}

The schema and API presented in
Sections~\ref{sec:design}-\ref{sec:apidesign} were reached through several DSR
build-evaluate cycles rather than in a single step. Evaluation included
comparison against QProv, mapping publicly documented provider interfaces,
constructing example provenance records, and implementing the fixture-backed
reference prototype. Table~\ref{tab:dsriterations} summarizes how gaps
identified during these activities led to refinements of the artifact.

\begin{table}[!htbp]
\centering
\caption{Design refinements of the provenance schema and unified API.}
\label{tab:dsriterations}
\footnotesize
\setlength{\tabcolsep}{3.5pt}
\renewcommand{\arraystretch}{1.1}

\begin{tabularx}{\textwidth}{c X X l}
\toprule
\textbf{It.}
& \textbf{Gap surfaced during evaluation}
& \textbf{Design decision}
& \textbf{Obj.} \\
\midrule

1
& QProv models quantum provenance but does not define the interface through
which the provenance was retrieved.
& Adopt QProv as the conceptual foundation and add software and access context.
& O1, O3 \\

2
& No provider exposes all applicable provenance attributes, and an absent
value is ambiguous.
& Introduce explicit evidence and availability states for provider-supplied,
derived, application-captured, unavailable, inapplicable, and unverifiable
information.
& O1, O5 \\

3
& Non-gate-model technologies require attributes that do not fit a
gate-oriented schema.
& Add optional, namespaced extensions while keeping a stable common core.
& O2, O4 \\

4
& Provenance is fragmented across job, device, characterization, program, and
result resources.
& Define a provider-independent provenance-record operation that aggregates and
normalizes the available resources.
& O2, O3 \\

5
& Coordinated adoption of one new provider API is unlikely in the short term.
& Introduce provider adapters that isolate retrieval and normalization from the
QMill-facing contract.
& O5 \\

6
& An aggregation platform and the hardware operator may be different entities.
& Represent the execution platform and underlying hardware provider separately.
& O2, O4 \\

7
& Providers return semantically different result representations, such as
observed counts and probability histograms.
& Preserve the provider-native result semantics and expose transformed values
only as explicitly derived representations.
& O1, O4 \\

\bottomrule
\end{tabularx}
\end{table}

The implementation cycle also refined the evaluation method. Raw counts of
evidence objects were initially found to be insufficient because one evidence
object does not necessarily correspond to one evaluated provenance attribute.
The final prototype therefore uses a fixed, versioned inventory of 21
attributes and classifies every inventory item independently for each
demonstration case. This prevents attributes from being silently omitted from
the coverage assessment.

\subsection{Answering the research questions}
\label{subsec:answers-rqs}

\noindent\textbf{\textit{RQ1: What provenance attributes are essential for
quantum algorithm development, execution tracking, and reproducibility?}}

The analysis identified six interrelated categories of essential provenance:
program and circuit artifacts, quantum-computer identity and characterization,
compilation, job and execution history, results, and software and access
context. These categories extend the QProv model~\cite{weder2021qprov} with
workflow-level information needed by applications that execute algorithms
across providers and over extended periods.

The program category must distinguish the original source, submitted,
compiled, and executed representations rather than treating a retrieved
provider program as the original source. Device provenance includes both
stable identity and time-dependent characterization. Compilation provenance
includes compiler and transpiler versions, optimization settings, mappings,
and generated artifacts. Execution provenance includes canonical and
provider-native job identifiers, status, shots, timestamps, queue duration,
execution duration, and failure information. Result provenance includes the
native result representation and any transformations applied to it. Software
and access context records the SDK, API, compiler, runtime, adapter, and
retrieval environment.

The evaluation also showed that the origin and availability of an attribute
are themselves essential provenance. A value returned by a provider, a value
derived from timestamps, and a value captured by the submitting application
must not be presented as equivalent evidence. Similarly, an unavailable value
must be distinguished from one that is not applicable to a provider or
technology. The schema therefore records evidence qualifiers in addition to
the attribute values.

These extensions address workflow-level experiment-tracking gaps discussed
in~\cite{gamage2025enhancingquantumsoftwaredevelopment} and support the need
for explicit execution context in NISQ-era
reproducibility~\cite{Preskill_2018}.

\medskip
\noindent\textbf{\textit{RQ2: To what extent do current quantum providers
expose essential provenance data, and through what mechanisms?}}

The comparative analysis covered 15 platforms: 12 hardware-provider platforms
and three cloud aggregators. It found substantial variation both in provenance
coverage and in the interfaces through which data are exposed. According to
the three-level classification used in the coverage matrix, execution
provenance had the strongest coverage: 87\% of the platforms were classified
as comprehensive and 13\% as partial. Compilation provenance was the weakest:
only one platform (Quantinuum) was classified as comprehensive, and the
remaining platforms were classified as partial. No platform comprehensively
exposed every evaluated provenance category.

The principal access mechanisms were provider SDKs, REST APIs, generated
clients, gRPC services, and cloud-platform resource models. Although most
platforms expose HTTP-based service interfaces, only five of the 15 provided
a public OpenAPI specification during the data-collection period. Other
platforms exposed machine-readable operations only through SDKs or generated
client models. Even when programmatic access existed, provenance was commonly
distributed across several resources rather than returned as one record.

Direct provider platforms generally documented richer access to
hardware-specific characterization and compilation context than cloud
aggregators. Aggregators provided comparatively consistent job identity,
status, and result retrieval, but hardware-specific provenance depended on
what the underlying provider exposed through the aggregation relationship.
The Braket-IonQ prototype case illustrates this distinction: Amazon Braket
supplies the task and platform model, while the hardware operator must be
retained separately and some provider-specific context remains unavailable.

These results characterize documented programmatic support at the July 2026
verification cutoff rather than field population in deployed provider
services (Section~\ref{subsec:study-design}). The study period itself
demonstrated the volatility of the quantum cloud ecosystem: interfaces were
retired, replaced, renamed, or expanded during data collection
(Section~\ref{subsec:provdataexposed}). Provider coverage must therefore be
treated as a time-bound property of a specific interface and version rather
than as a permanent property of the provider.

\medskip
\noindent\textbf{\textit{RQ3: What design principles should guide the
development of a unified quantum provenance API?}}

Section~\ref{subsec:designprinciples} identified the principal design
requirements: a standardized common schema, technology extensibility, an
HTTP-compatible interface, a machine-readable specification, time-aware
characterization, explicit availability metadata, and versioned backward
compatibility. The provider analysis and prototype refine these requirements
into six practical principles: normalization must preserve provider-native
semantics, such as the distinction between observed counts and returned
probabilities, rather than force all providers into the lowest common
representation; the schema must record the origin and strength of evidence,
including provider-supplied, aggregator pass-through, derived, and
application-captured values; the execution platform must be distinguishable
from the operator of the physical device; time-dependent characterization
must record the strength of its association with the job; application-side
capture is necessary for compilation and software information that provider
interfaces do not retain; and provider-specific retrieval and mapping should
be isolated behind adapters so that changes do not propagate into consuming
applications.

These principles combine established API-design
practices~\cite{lauret2019design} with requirements arising specifically from
heterogeneous quantum providers. They also explain why a unified provenance
API cannot be reduced to a common set of unqualified job fields: the
interface must normalize structure while retaining differences in meaning,
origin, and availability.

\medskip
\noindent\textbf{\textit{RQ4: Can a unified, evidence-aware provenance
interface be realized over existing heterogeneous provider interfaces in a
real-world quantum software architecture, and how does it handle incomplete
data and provider-interface change?}}

The QMill demonstration confirmed that provenance fragmentation introduces a
separate integration concern in addition to quantum-program execution.
Provider-specific implementations must retrieve and reconcile job, device,
characterization, program, result, and software information and map these
resources into application-level concepts. QMill's engineering experience
indicates that each provider integration carries a substantial implementation
and maintenance burden covering both execution and provenance
responsibilities.

The reference prototype demonstrates that the provenance portion of this
burden can be isolated behind a common contract: provider-specific retrieval
and normalization remain necessary
(Table~\ref{tab:prototypeimplementationresults}), but the application
accesses them through one common operation and contract, records degrade
gracefully when provenance is incomplete, and a controlled provider-interface
change was contained within the affected adapter
(Section~\ref{subsec:prototype-interpretation}).

Standardization therefore does not eliminate provider-specific complexity.
Its demonstrated benefit is to place that complexity behind a stable,
independently testable application boundary while making incomplete,
application-captured, and derived information explicit. The study does not
establish quantitative reductions in integration code or maintenance effort
(Section~\ref{sec:threats}).

\subsection{Interpretation of the QMill demonstration}
\label{subsec:prototype-interpretation}

The reference prototype provides evidence for four central properties of the
proposed design.

\noindent\textbf{\textit{Common-schema feasibility.}}
The Braket, IBM, and direct-IonQ records all validated against the same reduced
OpenAPI~3.1 contract, and no completed case contained an unclassified inventory
attribute. The common schema can therefore represent the selected
heterogeneous structures without exposing provider-specific fields in the
QMill-facing workflow.

This does not mean that the provider cases expose identical provenance.
Table~\ref{tab:prototypecoverage} shows distinct combinations of
provider-supplied, pass-through, derived, application-captured, unavailable,
and inapplicable information. The significance of the common contract is that
these differences remain explicit within one record structure.

\noindent\textbf{\textit{Semantic preservation.}}
The prototype avoids creating apparent uniformity by changing the meaning of
provider data. Braket and IBM return observed measurement counts, whereas the
IonQ case returns probabilities. The IonQ probabilities are preserved as the
provider-supplied canonical result representation. Count estimates are stored
separately as derived values and are not presented as observed counts.

Characterization is treated similarly. Braket characterization is classified
as current at retrieval, IBM characterization is the nearest available
provider snapshot with its time difference reported, and IonQ characterization
is provider-linked. These distinctions are necessary because all three
records may contain characterization metrics while offering different levels
of evidence that the metrics describe the device state relevant to the
execution.

\noindent\textbf{\textit{Combination of provider and application evidence.}}
The Braket and IBM demonstration records include deterministic
application-captured source, compilation, mapping, and software information.
The evidence qualifiers prevent these values from being confused with
provider-returned data. This confirms the finding from the provider comparison
that a complete provenance record cannot generally be assembled from provider
interfaces alone.

\noindent\textbf{\textit{Graceful degradation and change isolation.}}
The partial-record experiment (Section~\ref{subsec:prototype-results}) supports 
partial records as an honest representation of provider limitations rather than 
as failed complete records: no replacement snapshot was fabricated and the gap remained explicit.
The provider-input experiment similarly supports the role of adapters as compatibility 
boundaries, although one controlled field change cannot establish long-term maintenance savings.
The additive schema experiment addressed evolution of the common contract
rather than variation in provider input. Explicit versioning and strict validation can
therefore support controlled additive evolution.

\subsection{Implications for the quantum ecosystem}
\label{subsec:ecosystem-implications}

\noindent\textbf{\textit{For hardware providers.}}
Standardized, machine-readable provenance interfaces could reduce integration
friction and make provider capabilities easier to use from experiment
management, benchmarking, and auditing tools. The comparison suggests that
hardware providers can improve developer experience by linking jobs directly
to submitted programs, characterization identifiers, compilation settings, and
software versions. Even partial adoption would be useful if unavailable fields
and temporal associations were reported explicitly.

\noindent\textbf{\textit{For application developers.}}
A provider-independent provenance contract can reduce the amount of
provider-specific retrieval logic in applications and make cross-provider
analysis more systematic. Potential uses include experiment dashboards,
reproducible notebooks, benchmark archives, temporal device-performance
analysis, and calibration-aware compilation workflows. These applications
still need to interpret evidence qualifiers; normalization does not make
provider data equally complete or equally reliable.

\noindent\textbf{\textit{For cloud aggregators.}}
Aggregators provide important orchestration-level provenance, including common
job tracking, target identity, status, result retrieval, billing context, and
quota information. However, the comparative analysis shows that
hardware-specific characterization and compilation provenance may be less
complete than through direct provider interfaces. This limitation follows both
from differences among the underlying providers and from the need to map them
into a common platform model.

A standardized provenance model could allow aggregators to retain greater
granularity without requiring applications to understand every provider
interface. In particular, an aggregator could distinguish normalized platform
fields, underlying-provider pass-through fields, derived values, and unavailable
attributes rather than reducing all providers to an unqualified common
denominator. The Braket-IonQ case demonstrates how platform and hardware
identity can be preserved separately within such a model.

\noindent\textbf{\textit{For research communities.}}
Standardized provenance would facilitate reproducibility studies,
cross-provider benchmarking, and longitudinal analyses of quantum-hardware
performance. Explicit characterization timestamps and association strengths
would be particularly important when comparing executions performed at
different times. Machine-readable evidence origin would also allow researchers
to distinguish measured provider data from derived or application-captured
context.

\noindent\textbf{\textit{For software-engineering research.}}
Quantum computing exhibits interoperability challenges also found in other
emerging software ecosystems, including IoT, edge computing, and machine
learning. The results illustrate how an API-first contract and adapter-based
compatibility layer can provide a stable application boundary over
heterogeneous and evolving services, echoing established findings on
resource-oriented interface design~\cite{fielding2000rest} and on managing
API evolution, where versioning discipline and explicit compatibility
policies determine how interface changes propagate to
consumers~\cite{lubke2019apievolution}. Empirical studies of quantum-platform
defects similarly indicate that platform- and SDK-boundary changes are a
recurring source of quantum software faults, reinforcing the value of
isolating provider-specific logic behind stable
contracts~\cite{paltenghi2022bugs}. Quantum provenance adds the further
requirement that normalization preserve the evidential strength and temporal
relevance of the underlying data.

\subsection{Lessons from related domains}
\label{subsec:related-domain-lessons}

The machine-learning community faced related interoperability and
reproducibility challenges before the adoption of formats such as ONNX
\cite{onnx} and experiment-management systems such as
MLflow~\cite{zaharia2018mlflow}. ONNX illustrates the value of a shared,
versioned representation, while MLflow demonstrates the usefulness of
recording execution parameters, artifacts, metrics, and software context
together.

Containerization standards such as the Open Container Initiative\footnote{\url{https://opencontainers.org/}} address a
different layer of reproducibility by standardizing packaged software
environments. Quantum experiments require an analogous software context but
also depend on remote, time-varying hardware that cannot be fully captured in a
container image. Quantum provenance must therefore combine software
environment records with provider jobs, device characterization, compilation
context, and result artifacts.

W3C PROV~\cite{provdm2013} provides a relevant governance precedent: stable core
concepts can coexist with domain-specific extensions. A similar approach could
allow a stable quantum-provenance core to evolve alongside technology-specific
namespaces. Interest in reducing interface fragmentation across quantum-software 
ecosystems is also visible in the practitioner community~\cite{unitary}.

These precedents suggest that successful standardization requires more than a
technically adequate schema. It also requires transparent versioning,
conformance examples, governance, and adoption by providers, aggregators, and
software-tool developers. The adapter approach proposed in this work provides
an incremental path: applications can benefit from a common interface before
all providers adopt it natively.

\subsection{Threats to validity}
\label{sec:threats}

The threats to validity are discussed using the categorization of
Wohlin et al.~\cite{wohlin2012experimentation} in the context of applied research.

\noindent\textbf{\textit{Construct validity.}}
Provider terminology is not always semantically consistent. Terms such as
error rate, fidelity, execution time, compiled circuit, and calibration may
refer to different concepts or aggregation levels. A field may also be
documented without being populated for every job or device. These risks were
mitigated by using QProv as a conceptual reference, retaining provider-native
metric names where appropriate, distinguishing error from fidelity, and
defining explicit evidence and characterization-association states.

The fixed 21-attribute prototype inventory operationalizes only a subset of the
complete conceptual schema. Its classifications should therefore not be
interpreted as complete provider coverage. The inventory and the broader
provider-comparison matrix serve different purposes: the former evaluates the
implemented prototype, whereas the latter evaluates the documented provider
landscape.

\noindent\textbf{\textit{Internal validity.}}
The selection of provenance attributes and platforms may influence the observed
coverage. Essential attributes may have been omitted, or selected attributes
may favor providers with richer public documentation. These risks were reduced
through the systematic use of QProv, predefined inclusion criteria, and the
inclusion of 12 hardware-provider platforms and three aggregators across
superconducting, trapped-ion, neutral-atom, photonic, and
quantum-annealing technologies.

The prototype fixtures introduce an additional internal threat. They are
representative, sanitized structures rather than responses retrieved from live
authenticated services during the evaluation. Their construction may therefore
reflect interpretation errors or omit deployment-specific behavior. To reduce
this risk, the mappings were based on public SDK models, source code, schemas,
and examples; the fixtures are deterministic; all generated examples validate
against the contract; and automated drift checks keep the committed examples
and evaluation report synchronized with the implementation.

\noindent\textbf{\textit{External validity.}}
The provider ecosystem changes rapidly. Interfaces, field names, service
versions, hardware targets, and retention policies may differ after the
data-collection period. The results are therefore time-bound to the interfaces
and versions identified in the provider analysis. Recording the interface,
version, evidence source, and verification date improves repeatability but
does not eliminate this threat.

The study intentionally excludes authenticated provider testing. It does not
establish actual field population, permission behavior, retention, or
historical-data availability for deployed services. Similarly, the
fixture-backed prototype does not evaluate provider latency, reliability, rate
limits, or account- and region-specific behavior.

The principal prototype comparison uses Braket and IBM, with direct IonQ as a
supplementary case. Although these cases cover an aggregator and two direct
provider mappings, they do not establish generality across all providers or
hardware technologies. Further evaluation would be needed for neutral-atom, photonic,
and quantum-annealing interfaces.

\noindent\textbf{\textit{Conclusion validity.}}
The provider coverage matrix uses a qualitative three-level classification.
This simplifies complex interface capabilities and may obscure differences in
field completeness, accessibility, and semantic quality. The ratings were
mitigated through cross-checking of public documentation, SDK source code,
generated models, examples, and existing literature, together with detailed
provider descriptions that allow readers to inspect the basis of the
classification.

The prototype measurements also require careful interpretation. Source lines
are physical, non-blank, non-comment lines and do not directly measure
implementation complexity or engineering effort. The reported operations are
logical fixture-backed retrievals, not network calls. The provider-input
experiment contains one controlled field rename and cannot establish
long-term maintenance savings. 
The prototype consequently supports conclusions
about feasibility, explicit evidence handling, partial-record behavior, and
change localization. Quantitative claims about integration-effort or
maintenance-effort reductions are intentionally excluded: establishing them
would require an independently implemented provenance-only baseline and
longitudinal measurement of actual provider-interface changes and the
engineering effort of responding to them.

\section{Conclusions and future work}
\label{conclusions}

This paper addressed the absence of standardized, provider-independent access
to quantum execution provenance. The comparative analysis of 15 platforms
across five hardware technologies found substantial variation in provenance
coverage, terminology, and access mechanisms. No included platform
comprehensively exposed all evaluated provenance categories, and detailed
compilation provenance was consistently less complete than job, execution, and
result metadata.

Building on QProv~\cite{weder2021qprov}, we introduced an evidence-aware
provenance model that combines provider and application context without
conflating their origins. The model distinguishes artifact roles, provider
platforms and hardware operators, result representations, characterization
associations, derived values, unavailable data, and technology-specific
extensions.

Based on the provider analysis, we proposed a unified OpenAPI-based provenance
interface and provider-adapter architecture. The design normalizes record
structure while preserving differences in semantics, temporal relevance,
availability, and evidential strength. It provides an incremental path toward
standardization because existing provider interfaces can be mapped through
adapters without requiring coordinated native adoption.

The QMill reference prototype demonstrated the feasibility of this approach
for an aggregator-mediated Braket-IonQ execution, a direct IBM integration,
and a supplementary direct-IonQ case. The representative records validated
against one reduced contract while retaining their different evidence profiles
and result semantics. The controlled experiments additionally demonstrated
schema-valid partial records, localized response to a provider-input change,
and backward-compatible additive contract evolution.

The results do not show that provider-specific complexity can be eliminated.
Rather, they show that retrieval and normalization can be localized behind a
stable and independently testable application boundary. The demonstrated
benefits are common-contract feasibility, explicit treatment of incomplete
evidence, preservation of provider-specific meaning, and reduced propagation
of provider-interface changes into consuming applications.

The evaluation is intentionally limited to public interface evidence and
deterministic sanitized fixtures. It does not establish live field population,
provider-service reliability, or quantitative reductions in long-term
integration and maintenance effort. Nevertheless, it provides a reproducible
foundation for provider-authorized validation, broader adapter coverage, and
community refinement of a standard quantum-provenance interface.

As quantum computing develops toward sustained multi-provider workflows,
provenance should be treated as a foundational software interface rather than
as application-specific metadata. A shared, evidence-aware contract could
support reproducible experiment records, systematic cross-provider comparison,
and third-party tools without requiring providers to expose identical data or
adopt identical internal service architectures.

\subsection*{Future work}

The work can be extended in the following directions.

\begin{itemize}

    \item \textbf{Community-driven standardization and governance.}
    The proposed contract should be discussed and refined with quantum hardware
    providers, cloud aggregators, application developers, and standards
    organizations. Future work should define governance for the common schema,
    evidence vocabulary, provider identifiers, characterization associations,
    semantic versioning, extension namespaces, and compatibility policies.
    Potential venues for such coordination include open-source communities
    working on unified quantum interfaces, such as the QIR
    Alliance~\cite{qir} and the Unitary Foundation\footnote{\url{https://unitaryfoundation.org}},
    as well as IEEE quantum-computing standards activities.

    \item \textbf{Provider-authorized operational validation.}
    The current provider comparison and prototype intentionally rely on public
    documentation, SDK models, schemas, examples, and sanitized fixtures.
    Future studies should validate the mappings using provider-authorized
    execution records. This would make it possible to evaluate actual field
    population, authentication scopes, job-data retention, historical
    characterization availability, regional variation, rate limits, and
    provider-service latency. Credential-free fixtures should remain the
    reproducible baseline even when live validation is added.

    \item \textbf{Broader provider and technology coverage.}
    The principal prototype comparison covers Amazon Braket and IBM Quantum,
    with direct IonQ as a supplementary mapping. Future adapters should cover
    additional direct providers and aggregators, including platforms
    representing neutral-atom, photonic, quantum-annealing, and other emerging
    architectures. Such work should evaluate whether the stable common core and
    namespaced extensions remain adequate across non-gate-model technologies.

    \item \textbf{Production API and persistence architecture.}
    The prototype exposes the proposed operations through an in-process service
    and does not implement a production HTTP deployment or durable provenance
    store. Future work should evaluate authentication and authorization,
    multi-tenant canonical identifiers, encrypted provider-native artifacts,
    artifact-retention policies, refresh semantics, caching, version migration,
    and access control for potentially sensitive provider metadata.

    \item \textbf{Artifact lineage and hybrid workflows.}
    Further work should extend artifact relationships among source, submitted,
    compiled, executed, and result representations. Particular attention is
    needed for hybrid quantum-classical workflows in which one experiment
    contains several quantum jobs, classical optimization steps, intermediate
    parameter updates, mitigation procedures, and repeated executions on
    different devices.

    \item \textbf{Time-dependent characterization and calibration drift.}
    Longitudinal studies should evaluate how device characterization changes
    relative to job execution and how strongly a provider can associate a
    characterization snapshot with a specific run. This includes defining
    matching thresholds for nearest-available snapshots, representing
    uncertainty and validity intervals, and evaluating calibration-aware
    compilation and result interpretation.

    \item \textbf{Result and metric semantics.}
    Additional work is needed to standardize the representation of raw shots,
    observed counts, probabilities, quasi-probabilities, mitigated results,
    expectation values, and technology-specific outputs. Controlled
    vocabularies and unit conventions are also needed for characterization
    metrics while preserving provider-native definitions and avoiding
    unjustified conversions between errors and fidelities.

    \item \textbf{Longitudinal maintenance and usability evaluation.}
    Future work should observe actual provider-interface changes and measure their
    effects on adapters, the common implementation, and consuming applications.
    User studies could additionally evaluate whether evidence-aware provenance
    improves debugging, comparison, reproducibility, and confidence in experimental
    results.

    \item \textbf{Conformance tools and reusable clients.}
    The implemented contract, inventory, examples, and drift checks provide a
    foundation for provider-adapter conformance suites, schema-linting tools,
    compatibility reports, and generated clients. Such tools
    should validate not only structural conformance but also evidence origin,
    temporal association, units, artifact lineage, and explicit treatment of
    unavailable data.

\end{itemize}

\section*{Acknowledgments}
This work has been supported by Business Finland through project EM4QS (155/31/2024), and the Research Council of Finland through Finnish Quantum Flagship (359240, JYU) \& project DEQSE (349945). We thank QMill Oy
for providing the industrial context and access to its multi-provider
quantum-software architecture for the reference-prototype demonstration.

\bibliography{references}

\end{document}